\documentclass{aa}
\bibpunct{(}{)}{;}{a}{}{,} % to follow the A&A style
\usepackage{graphicx}
\usepackage{multirow}
\usepackage{booktabs}
\usepackage{txfonts}
\usepackage{color}
\usepackage[normalem]{ulem}

\begin{document}

   \title{Testing the spatial geometry of the universe with TianQin: the prospect of using supermassive black hole binaries}

   \author{Yu Pan
          \inst{1,2}
          \and
         Jingwang Diao\inst{1}
          \and
          Jing-Zhao Qi\inst{3}
          \and
          Jin Li\inst{2,4}\fnmsep\thanks{corresponding author}
          \and
          Shuo Cao\inst{5,6}\fnmsep\thanks{corresponding author}
          \and
          Qing-Quan Jiang\inst{7}\fnmsep\thanks{corresponding author}
          }

   \institute{School of Science, Chongqing University of Posts and Telecommunications, Chongqing 400065, China\\
              \email{panyu@cqupt.edu.cn}
\and
Department of Physics and Chongqing Key Laboratory for Strongly Coupled Physics, Chongqing University, Chongqing 401331, China
         \and
             Department of Physics, College of Sciences, Northeastern University, Shenyang 110004, China
         \and
            Department of physics, Chongqing University, 400044 Chongqing, China\\
                \email{cqujinli1983@cqu.edu.cn}
         \and
            Institute for Frontiers in Astronomy and Astrophysics, Beijing Normal University, Beijing 102206, China\\
                \email{caoshuo@bnu.edu.cn}
         \and
         Department of Astronomy, Beijing Normal University, Beijing 100875, China
         \and
            School of Physics and Astronomy, China West Normal University, Nanchong 637009, China\\
                \email{qqjiangphys@yeah.net}
        }

   \date{Received xx; accepted xx}

% \abstract{}{}{}{}{}
% 5 {} token are mandatory

  \abstract
   {The determination of the spatial geometry of the universe plays an important role in modern cosmology. Any deviation from the cosmic curvature $\Omega_K=0$ would have a profound impact on the primordial inflation paradigm and fundamental physics.}
   {In this paper, we carry out a systematic study of the prospect of measuring cosmic curvature with the inspiral signal of supermassive black hole binaries (SMBHBs) that could be detected with TianQin.}
   {The study is based on a cosmological-model-independent method that extended the application of gravitational wave (GW) standard sirens in cosmology. By comparing the distances from future simulated GW events and \textbf{simulated $H(z)$ data}, we evaluate if TianQin would produce robust constraints on the cosmic curvature parameter $\Omega_{k}$. More specifically, we consider \textbf{3-yr} to 10-yr observations of supermassive black hole binaries with total masses ranging from $10^{3}M_\odot$ to $10^{7}M_\odot$.}
   {Our results show that in the future, with the synergy of 10-yr high-quality observations, we can tightly constrain the curvature parameter at the level of $1\sigma$ \textbf{$\Omega_k=-0.002\pm0.061$}.
   Moreover, our findings indicate that the total mass of SMBHB does influence the estimation of cosmic curvature, implied by the analysis performed on different subsamples of gravitational wave data.
   }
   {Therefore, TianQin is expected to provide a powerful and competitive probe of the spatial geometry of the universe, compared to future spaced-based detectors such as DECIGO.}

   \keywords{cosmic curvature --
                gravitational wave --
                Gaussian processes
               }

   \maketitle
%
%________________________________________________________________

\section{Introduction}
\label{sec1}

As a successful gravity theory where general coordinate invariance acts an essential role, Einstein's Theory of General Relativity (GR) has passed all observational test so far, ranging from the submillimeter scales \citep{Li2003,Li2008} to galactic scales \citep{Cao2017}. In recent years, the appearance of GWs has provided us with a new window to study cosmology. Prior to the detection of GWs, all cosmological parameter inferences were made using electromagnetic (EM) radiation from extragalactic sources. Currently, we can detect GWs produced by inspiraling binary systems that provide absolute distance information \citep{Schutz1986}. Therefore, we can use these binary systems as standard sirens. Gravitational waves have been applied in various ways in cosmology \citep{Zhao2011,Cai2017,Liao2019,Liao2019a,Liao2017a,Liao2022,He2022,Pan2021,Zhang2023}.

Unlike most of these works that pay attention to the effects of weak gravitational fields, gravitational waves detection supports a new approach to testing GR in the strong field. Fortunately, since the successful detection of the first GW event by the advanced LIGO ((GW150914 \citep{Abbott2016,Abbott2016a,Abbott2016b,Abbott2016c,Abbott2017,Abbott2017a,Abbott2017b})), several GW events from mergers of binary black holes \citep{Abbott2016,Abbott2016a} and one GW event accompanied by an electromagnetic counterpart from the merger of binary neutron stars have been observed \citep{Abbott2018,Abbott2020}. Furthermore, GWs are helpful for us to explore some extreme conditions, such as the very early universe, extremely high energy scales, extra dimensions and so on. Hence, the detection of GWs has received considerable attention \citep{Petiteau2011}. New generations of GW detectors have been proposed to cover different frequency bands. With these detectors, more binary coalescences are expected to be detected at longer distances, and they have higher a signal to noise ratio (SNR). These sources are very important for us to study the universe's evolution \citep{Petiteau2011}. At present, the gravitational wave detection projects promoted both at home and abroad include ground-based and space-based GW detectors. The ground-based GW detectors include LIGO and VIRGO observatories, KAGRA and ET GW detectors, etc. The space-based projects include the LISA, DECIGO, and two Chinese detectors, TianQin (TQ) and Taiji \citep{Gong2021}, etc. Space-based detectors can detect GWs below 1Hz, but ground-based detectors cannot, because they are affected by terrestrial gravity gradient noise. The TQ Project, first proposed by Sun Yat-sen University in China in 2014, is a space-based gravitational wave detector with some features. It is capable of detecting GWs in the range of $10^{-4}$ to $10^{-1}$ Hz to fill the gap between DECIGO and LISA. In addition, the equilateral triangle arm of TQ is around $10^5$ km. Because of the shorter arm and better sensitivity comparing with LISA in the higher frequency regime, it might perform better than LISA \citep{Gong2021}. There are three steps for TQ project. The successful launch of the TQ-1 satellite in 2019 has achieved all mission objectives, which is the main objective in the first step of the project. The main goal of the second step of the project is to launch two satellites in 2025. TQ-2 would be able to detect GWs if it didn't have a lot of laser phase noise affecting it. The third step of the project, which is expected to take place in 2035, aims to ensure that the three satellites operate smoothly in geocentric orbit and can successfully detect GWs, but this is subject to the successful implementation of the first two steps. The TQ GW detector can observe GW signals including SMBHBs, stellar-mass binary black hole mergers (SBBH), and extreme mass ratio inspirals (EMRIs) \citep{Zhu2022}.

Many studies have been done on SMBHBs. For example, \cite{Feng2019} used the Fisher information matrix to estimate the parameter precision of SMBHBs for TQ. They found that the parameters of SMBHBs could be determined with a high precision, which means that the SMBHBs are the most powerful GW sources detected by TQ. In addition, the universe's spatial curvature is one of the most important problems in cosmology. Its importance stems from the following three aspects. Firstly, the estimation of
cosmic curvature parameter could provide an important probe of the well-known Friedman -- Lema{\^i}tre -- Robertson -- Walker (FLRW) metric \citep{Cao2019,Qi2019}. Secondly, the cosmic curvature is closely related to the universe's evolution and the hypostasis of dark energy (DE). Even a very small curvature of the universe has a very important effect on the reconstruction of the DE equation of state \citep{Clarkson2007,Gong2007,Virey2008,Ichikawa2006}.
Moreover, the failure of the FLRW approximation might provide an explanation of late-time universe's accelerating expansion \citep{Ferrer2006,Enqvist2008,Ferrer2009,Rasanen2009,Boehm2013,Lavinto2013,Redlich2014}. Therefore, it is necessary to constrain the cosmic curvature from popular observational probes and indeed it has been extensively studied in the literature \citep{Zhao2007,Wright2007,Wei2017,Li2016}. In this work, we investigate the potential of constraining the cosmic curvature with SMBHBs covering the total mass range of $10^3-10^{7}M_\odot$.

Generally, three theoretical methods can be used to constrain the spatial curvature of the universe. One is the model-dependent method using the photometric distance expressed by the cosmological constant model, which does not directly measure the cosmic curvature geometrically. Another approach is the model-independent method based on the zero geodesic distance of the FLRW metric to constrain the spatial curvature \citep{Bernstein2006,Liao2017,Qi2018,Denissenya2018}. The third model-independent method was proposed to constrain $\Omega_k$ from the measurements of expansion rate ($H(z)$), the comoving distance ($D(z)$) and its derivative with respect to $z$ \citep{Clarkson2007}: $\Omega_{K} = \frac{(H(z)D^{\prime}(z))^2-c^2} {H_{0}^{2}D(z)^2}$. At present, many works have used this method to constrain the cosmic curvature based on the observations of different types of cosmological distances \citep{Shafieloo2010,Cao2017,Mortsell2011,Cai2016,Sapone2014,Li2014}.
However, one needs to estimate the first derivative of $D'(z)$, which may bring a large uncertainty to the final estimation of $\Omega_K$.
In this work, we study the potential of an improved cosmological-model-independent method that extended the application of gravitational wave standard sirens in cosmology, based on the inspiral signal of supermassive black hole binaries that could be detected with TQ. In the framework of such methodology, \cite{Wei2017} proposed the Gaussian process (GP) method to constrain the cosmic curvature by combining the most recent Hubble parameter $H(z)$ and supernova Ia data, which turned out to agree with zero cosmic curvature. By comparing the distances from future simulated GW events and current CC Hubble data, \cite{Wei2018} continued to expand the application of GW standard sirens in measuring cosmic curvature. They simulated hundreds of GW data from ET to constrain the curvature parameter and obtained the higher precision result, for which $\Omega_k=-0.002\pm0.028$. More recently, \cite{He2022} reconstructed the Hubble parameters without the influence of the hypothetical model, and combined with DECIGO's GW data to give a high-precision result ($\Omega_k=-0.07\pm0.016$).

The paper is organized as follows. The simulation of the TianQin data is introduced in Section \ref{sec.data}. The methodology and the corresponding constraints on cosmic curvature from the TQ are presented in Section \ref{sec.result}. The final discussion and conclusions are summarized in Section \ref{sec.conclusion}.

%__________________________________________________________________

\section{Observational data}\label{sec.data}

\begin{figure*}[htbp]
\begin{center}
\footnotesize
\begin{tabular}{ccc}
\includegraphics[width=0.5\textwidth]{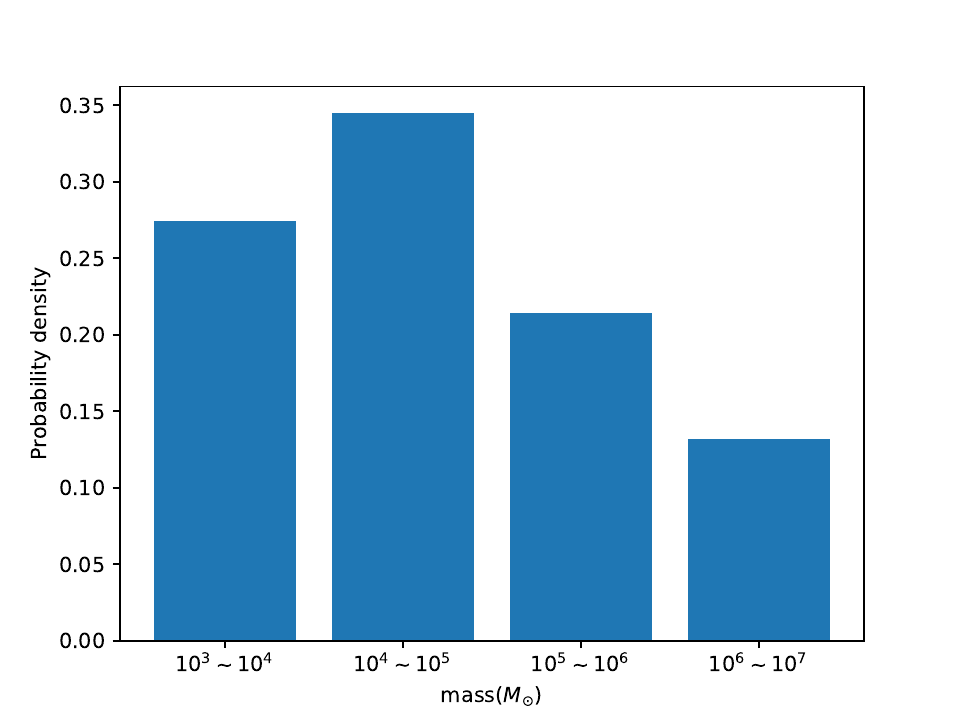} &
\includegraphics[width=0.5\textwidth]{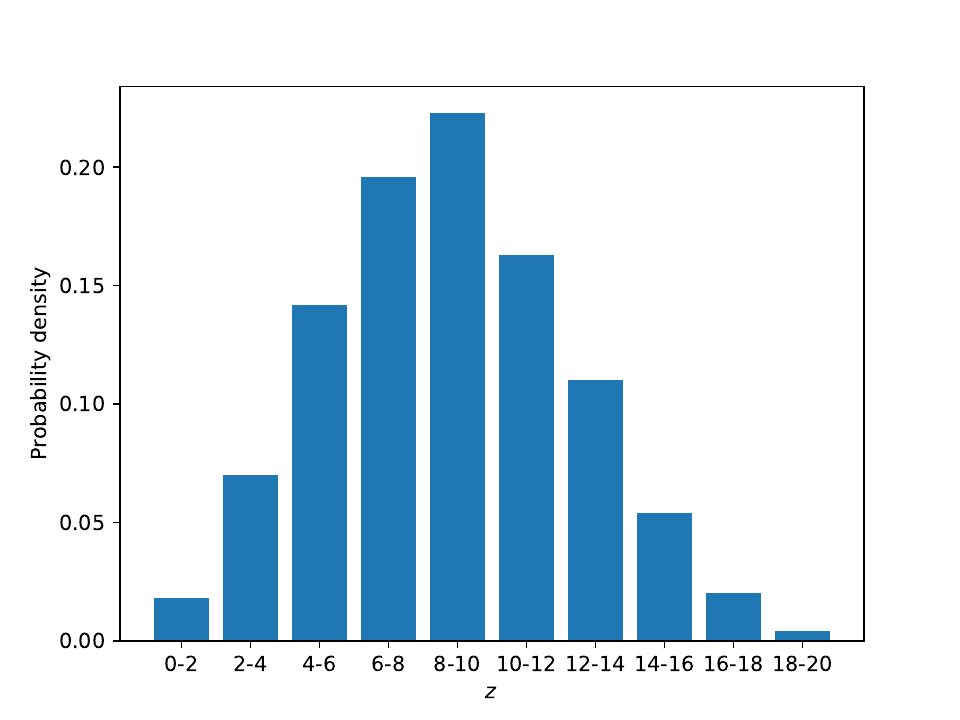} &\\
(a) & (b)\\
\end{tabular}
\end{center}
\caption{The event rates for different mass ranges (left panel) \textbf{and redshift ranges} (right panel) based on the popIII model.}
\label{possibilty}
\end{figure*}

\begin{table}
\begin{center}
\footnotesize
\caption{The $H(z)$ data measurement from the differential age method (I) and the radial BAO method (II).}
\begin{tabular}{cccl}
\toprule
$z$                   &~~$H(z)(km/s/Mpc)$  &~~ Method  &~~References~~   \\
\hline
$0.09$     &$69\pm 12$   &I  &\cite{Jimenez2003} \\
\hline
$0.17$     &$83\pm 8$      &I   &\multirow{8}*{\cite{Simon2005}} \\

$0.27$    &$77\pm 14$    &I   &\multirow{8}*{ }\\

$0.4$    &$95\pm 17$     &I  &\multirow{8}*{ }\\

$0.9$    &$117\pm 23$      &I  &\multirow{8}*{ }\\

$1.3$    &$168\pm 17$      &I  &\multirow{8}*{ }\\
$1.43$    &$177\pm 18$      &I  &\multirow{8}*{ }\\
$1.53$    &$140\pm 14$      &I  &\multirow{8}*{ }\\
$1.75$    &$202\pm 40$     &I   &\multirow{8}*{ }\\
\hline
$0.48$    &$97\pm 62$     &I   &\multirow{2}*{\cite{Stern2010}}\\
$0.88$    &$90\pm 40$     &I   &\multirow{2}*{ }\\
\hline
$0.35$    &$82.1\pm 4.9$     &I   &\cite{Chuang2012}\\
\hline

$0.179$    &$75\pm4$     &I   &\multirow{8}*{\cite{Moresco2012}}\\
$0.199$    &$75\pm5$     &I   &\multirow{8}*{ }\\
$0.352$    &$83\pm14$     &I   &\multirow{8}*{ }\\
$0.593$    &$104\pm 13$     &I   &\multirow{8}*{ }\\
$0.68$    &$92\pm 8$     &I   &\multirow{8}*{ }\\
$0.781$    &$105\pm 12$     &I   &\multirow{8}*{ }\\
$0.875$    &$125\pm 17$     &I   &\multirow{8}*{ }\\
$1.037$    &$154\pm 20$     &I   &\multirow{8}*{ }\\
\hline
$0.07$    &$69\pm 19.6$     &I   &\multirow{4}*{\cite{Zhang2014}}\\
$0.12$    &$68.6\pm 26.2$     &I   &\multirow{4}*{ }\\
$0.2$    &$72.9\pm 29.6$     &I   &\multirow{4}*{ }\\
$0.28$    &$88.8\pm 36.6$     &I   &\multirow{4}*{ }\\
\hline
$1.363$    &$160\pm33.6$     &I   &\multirow{2}*{\cite{Moresco2015}}\\
$1.965$    &$186.5\pm 50.4$   &I   &\multirow{2}*{ }\\
\hline
$0.3802$    &$83\pm 13.5$    &I    &\multirow{5}*{\cite{Moresco2016}}\\
$0.4004$    &$77\pm 10.2$     &I   &\multirow{5}*{ }\\
$0.4247$    &$87.1\pm 11.2$    &I    &\multirow{5}*{ }\\
$0.4497$    &$92.8\pm 12.9$     &I   &\multirow{5}*{ }\\
$0.4783$    &$80.9\pm 9$    &I    &\multirow{5}*{ }\\
\hline
$0.24$    &$79.69\pm 2.65$     &II   &\multirow{2}*{\cite{Gaztanaga2009}}\\
$0.43$    &$86.45\pm 3.68$  &II  &\multirow{2}*{ }\\
\hline
$0.44$    &$82.6\pm 7.8$  &II  &\multirow{3}*{\cite{Blake2012}}\\
$0.6$    &$87.9\pm 6.1$  &II  &\multirow{3}*{ }\\
$0.73$    &$97.3\pm 7$  &II  &\multirow{3}*{ }\\
\hline
$0.35$    &$84.4\pm 7$  &II  &\cite{Xu2013}\\
\hline
$0.57$    &$92.4\pm 4.5$  &II  &\cite{Samushia2013}\\
\hline
$2.3$    &$224\pm 8$  &II  &\cite{Busca2013}\\
\hline
$2.36$    &$226\pm 8$  &II  &\cite{FontRibera2014}\\
\hline
$2.34$    &$222\pm 7$  &II  &\cite{Delubac2015}\\

\bottomrule
\end{tabular}
\label{hzdata}
\end{center}
\end{table}

In this section, we briefly introduce the method employed to simulate observations of GW standard sirens from the TQ. In the following simulations, we adopt the flat $\Lambda$CDM with the Hubble constant $H_0=69.6$ km/s/Mpc with 1\% uncertainty and the matter density parameter $\Omega_m=0.286$ \citep{Bennett2014}.

\begin{figure*}[htbp]
\begin{center}
\footnotesize
\begin{tabular}{ccc}
\includegraphics[width=0.5\textwidth]{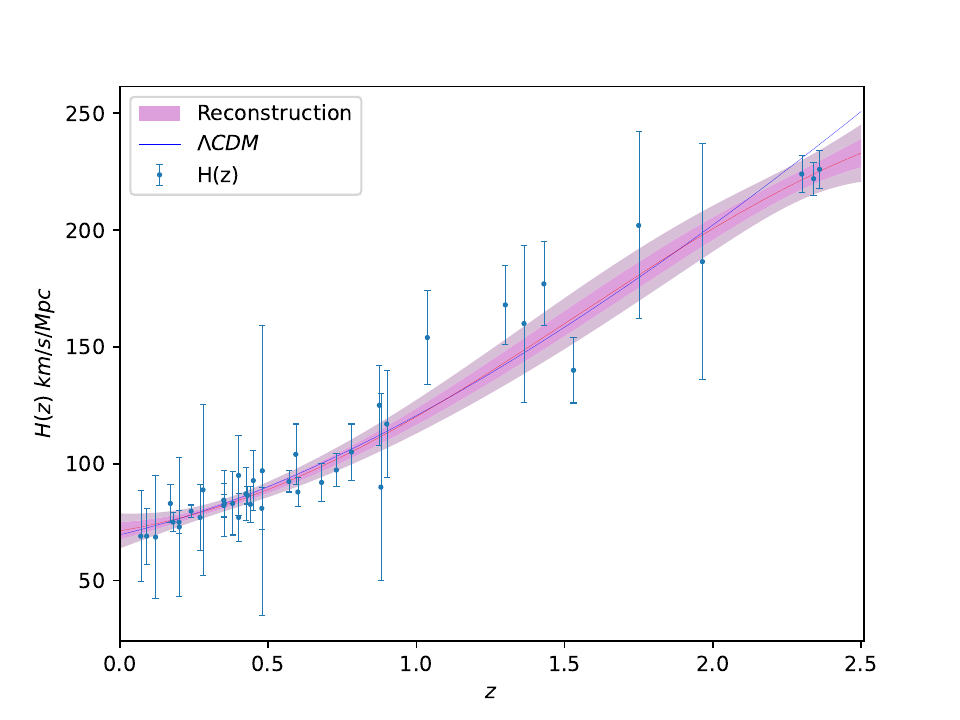} &
\includegraphics[width=0.5\textwidth]{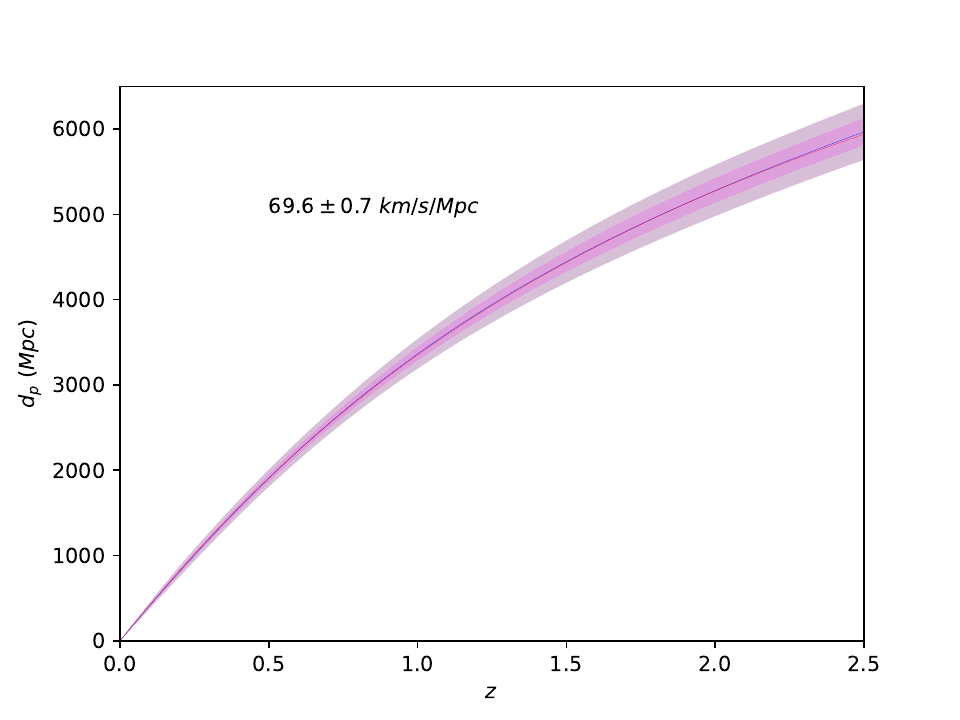} &\\
(a) & (b)\\
\end{tabular}
\end{center}
\caption{The reconstructed $H(z)$ and $d_p(z)$ function by GP method using the $H(z)$ data with $H_0=69.6\pm0.7$ km/s/Mpc. The $H(z)$ observational data and the $\Lambda$CDM model is also added for comparison. }
\label{Ezdp_real}
\end{figure*}

\begin{figure*}[htbp]
\begin{center}
\footnotesize
\begin{tabular}{ccc}
\includegraphics[width=0.5\textwidth]{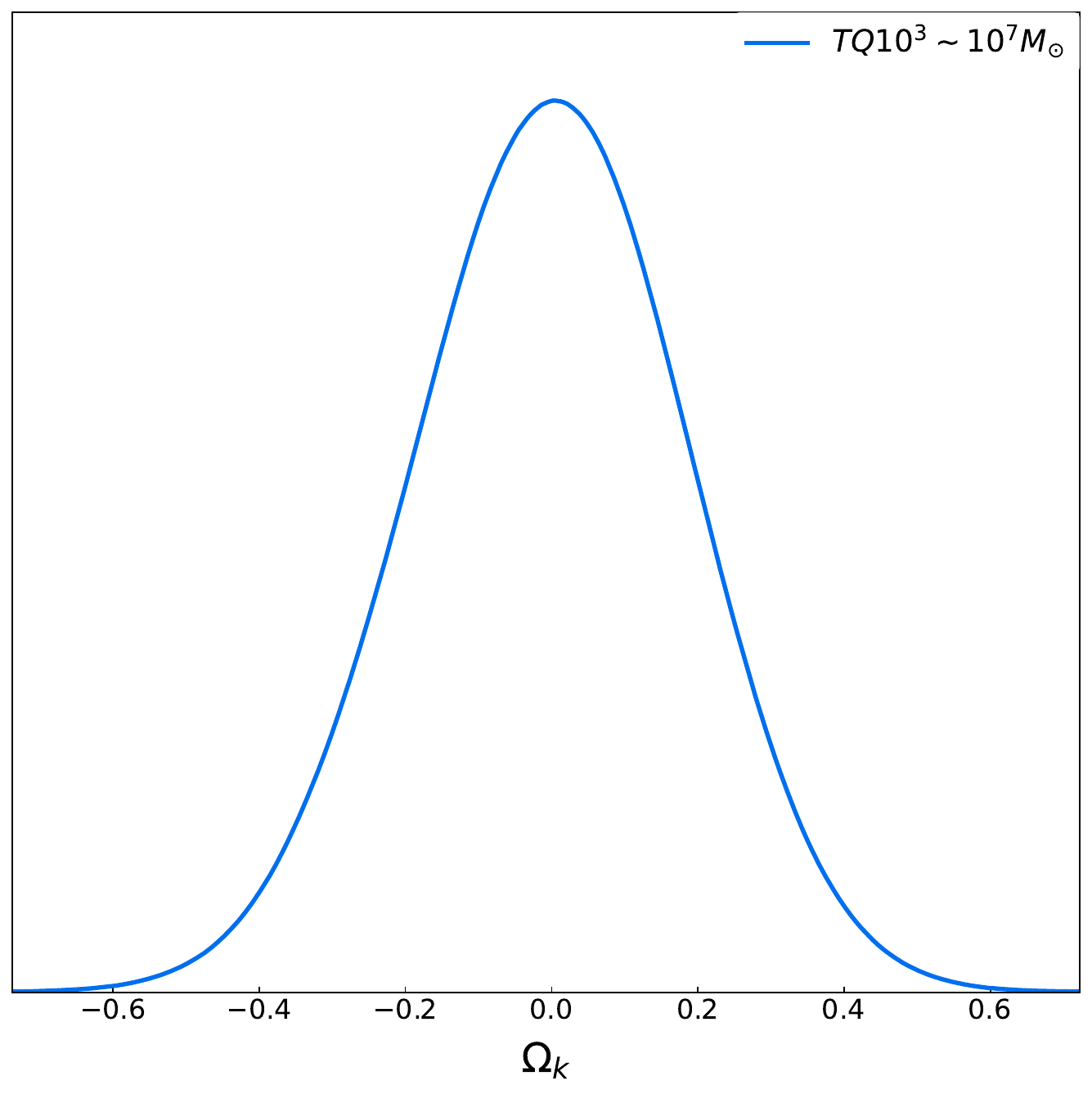} &
\includegraphics[width=0.5\textwidth]{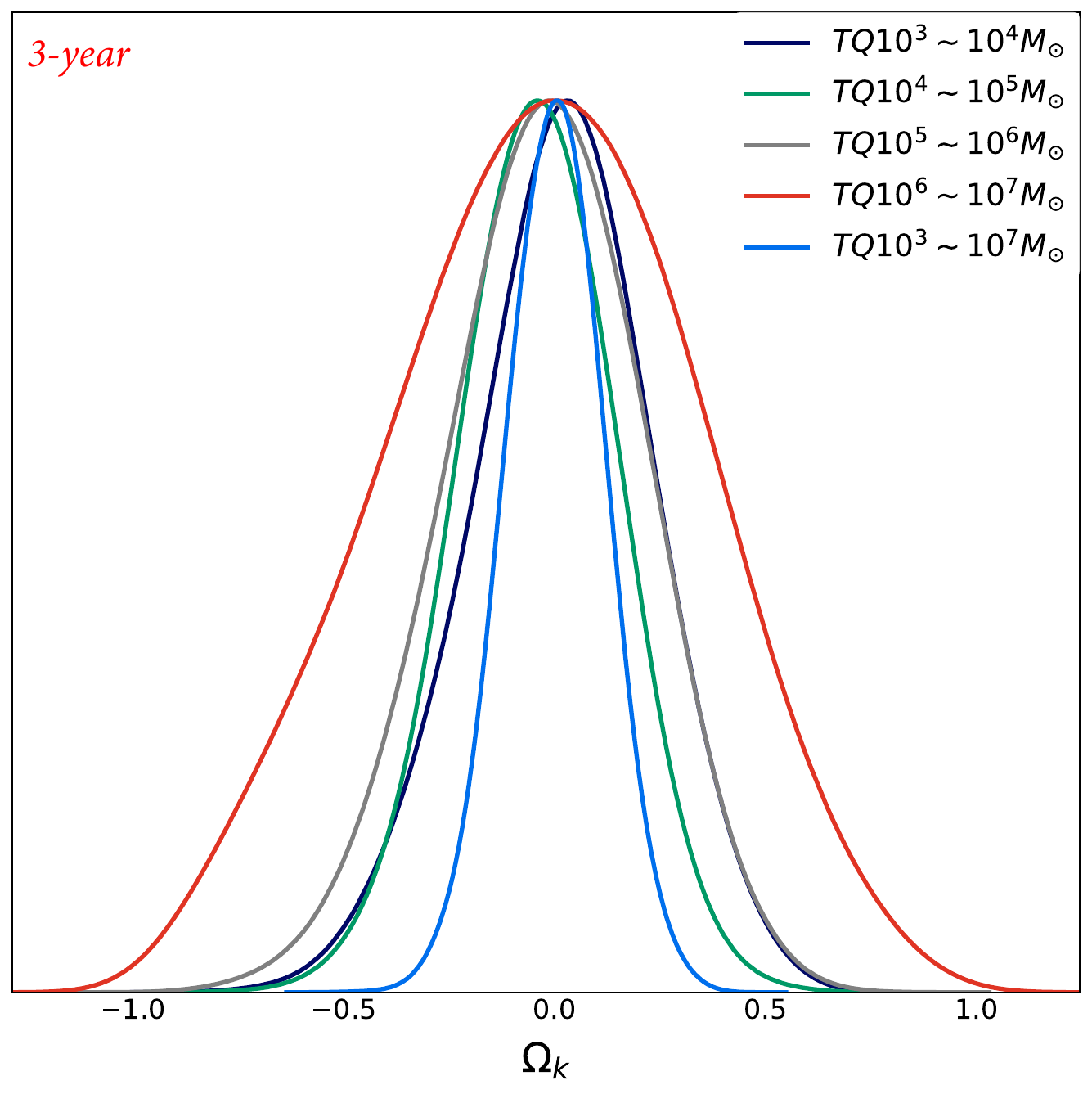} &\\
(a) & (b)\\
\includegraphics[width=0.5\textwidth]{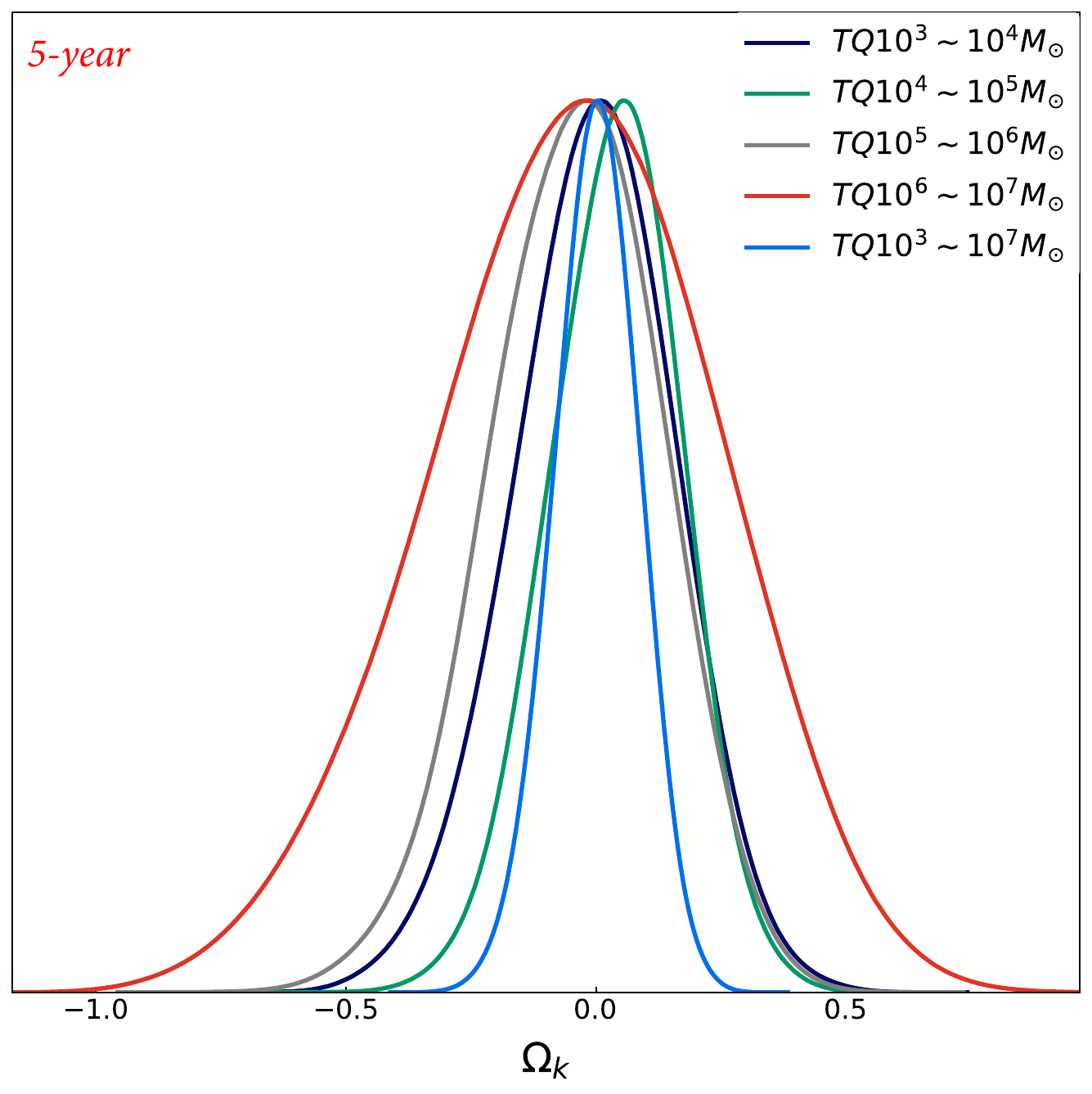} &
\includegraphics[width=0.5\textwidth]{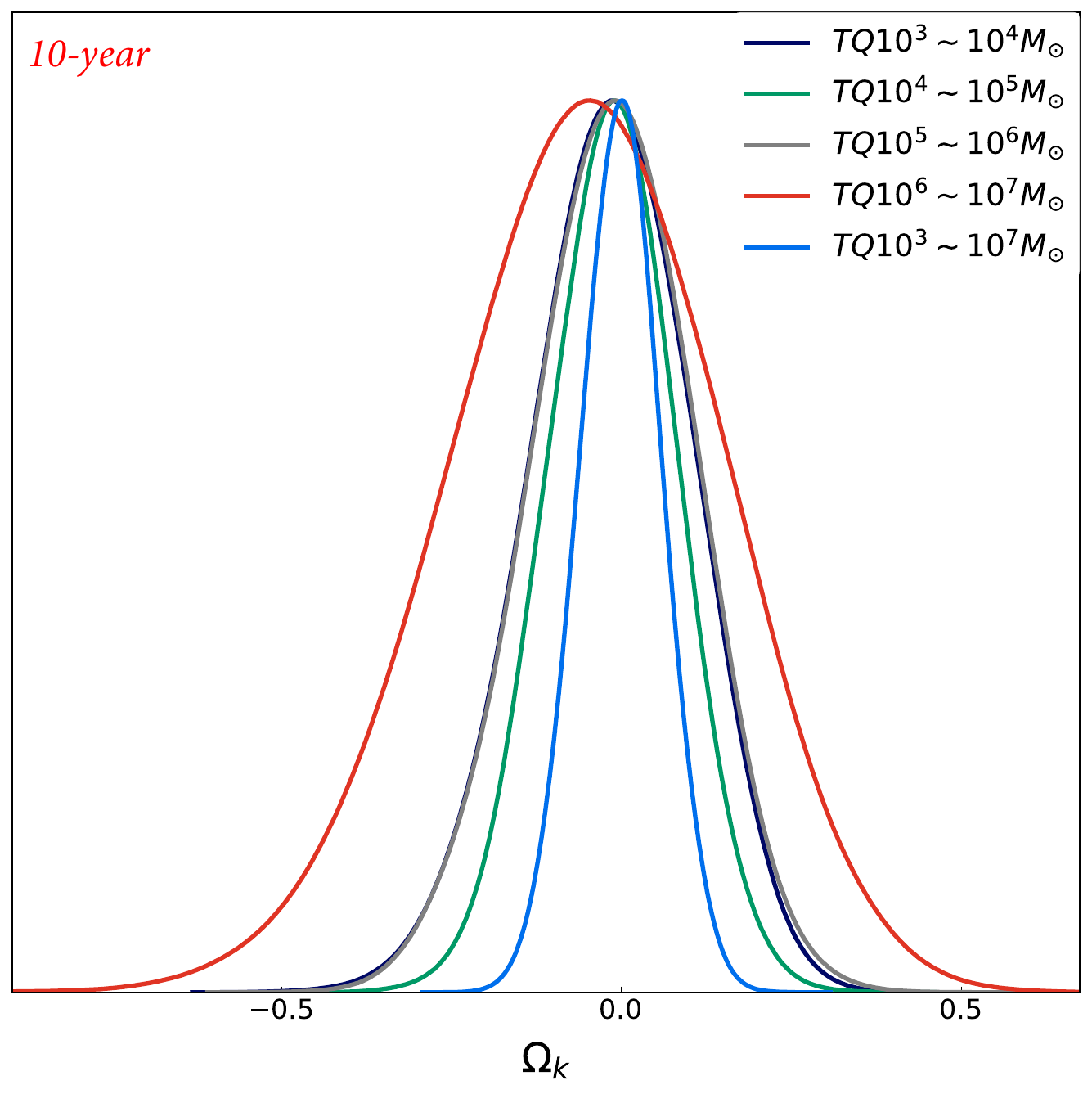} &\\
(c) & (d)\\
\end{tabular}
\end{center}
\caption{The 1D distribution of cosmic curvature is derived from different sub-samples of SMBHBs. Subplot (a) represents the result of $H(z)$ data combined with 10-year observations of TQ, while subplots (b), (c) and (d) correspond to the results of simulated $H(z)$ data combined with 3-, 5-, and 10-year observations of TQ, respectively.}
\label{TianQinP}
\end{figure*}

As a typical millihertz frequency gravitational wave observatory, TQ is designed to mainly detect GWs from Galactic ultra-compact binaries, coalescing SMBHBs, EMRIs, and inspiral of SBBH \citep{Hughes2001,Sesana2016,Di2018,Olmez2010}. The SMBHB mergers are considered to be the most powerful GW sources that could be observed by TQ \citep{Feng2019}. Therefore, we simulated GW datasets of SMBHBs for 3-, 5- and 10-year, respectively, according to the continuous observation time of TQ in the future. For each year's data which have masses ranging from $10^3M_\odot$ to $10^7M_\odot$, and split the datasets into 3 sub-samples according to the mass range of SMBHBs (named TQ$10^{3}\sim 10^{4}M_\odot$, TQ$10^{4}\sim 10^{5}M_\odot$, TQ$10^{5}\sim 10^{6}M_\odot$ and TQ$10^{6}\sim 10^{7}M_\odot$, respectively). For each sub-sample, the redshift $z$ is chosen in the range of $0\sim2.5$. Moreover, for the event rate of SMBHBs that Tianqin can detect each year, we refer to the recent results based on the popIII model, which assumes that light massive black hole seeds from popIII stars and accounts for the delays between massive black hole and galaxy mergers \citep{Yi2022}. In Fig.~\ref{possibilty}, we present the event rates for different mass intervals \textbf{and redshift intervals}, the distribution of which agrees well with that of \citet{Klein2016}. It should be noted that since TQ has two observation windows of two and three months each year \citep{Luo2016}, the practically detected SMBHB event is only half of the predicted number proposed in \citet{Yi2022}.

It is well known that the absolute measurements of the luminosity distance $D_L$ to the source and the chirp mass $M_c$ could be determined from the chirping signals of GW \citep{Schutz1986}. In addition, one can measure the chirp mass by the phase of the GW signal, hence one can extract $D_{L}$ from the strain amplitude of GW. For the waveform of GW, we choose $\mathcal{H}(f)$ in stationary phase approximation.
\begin{align}
\mathcal{H}(f)=Bf^{-7/6}\exp[i(2\pi ft_0-\pi/4+2\psi(f/2)-\varphi_{(2.0)})],
\label{equa:hf}
\end{align}
where $B$ is the Fourier amplitude and
\begin{align}
B=&~~\frac{c}{D_L}\sqrt{F_+^2(1+\cos^2(\iota))^2+4F_\times^2\cos^2(\iota)}\nonumber\\
            &~~\times \sqrt{5\pi/96}\pi^{-7/6}\mathcal{M}_c^{5/6},
\label{equa:A}
\end{align}
where $M_c=\frac{(1+z)(m_{1}m_{2})^{3/5}}{(m_{1}+m_{2})^{1/5}}$ is the chirp mass. See \cite{Feng2019} for the expression of $F_{+}$, $F_{\times}$ and phase parameters for advanced TQ. In addition, $D_{L}$ is the luminosity distance and in the standard flat $\Lambda$CDM cosmological model it can be expressed as
\begin{equation}
 D_{L}=\frac{c(1+z)}{H_{0}}\int^{z}_{0}\frac{dz'}{\sqrt{\Omega_{m}(1+z)^{3}+1-\Omega_{m}}},
\end{equation}
The one-sided power spectral density (PSD) of the noise in TQ is \citep{Feng2019}
\begin{equation}
S_{n}(f)=\frac{S_{x}}{L^{2}}+\frac{4S_{a}}{(2\pi f)^{4}L^{2}}(1+\frac{10^{-4}Hz}{f}),
\end{equation}
where $S_{x}=10^{-24}$m$^{2}/$ Hz denotes the PSDs of the position noise, $L=1.73\times10^{5}$ km denotes the arm length, and $S_{a}=10^{-30}$m$^{2}$s$^{-4}/$ Hz denotes the PSDs of residual acceleration noise, respectively.	
Correspondingly, the SNR of TQ can be calculated as
\begin{align}
\rho=\frac{(M_{c}G)^{5 / 6}}{\sqrt{10} \pi^{2 / 3} D_{L}c^{3/2}} \sqrt{\int_{f_{\text {in }}}^{f_{\text {fin }}} \frac{f^{-7 / 3}}{S_{n}(f)} \mathrm{d} f},
\end{align}
where $f_{fin} = min(f_{ISCO}, f_{end})$ and the $f_{ISCO} =c^{3}/(6^{3/2}MG\pi)$ Hz is the GW frequency at the innermost stable circular orbit, $f_{end} = 1$ Hz is the upper cutoff frequency for TQ, $f_{in} = max(f_{low}, f_{obs})$ with the lower cutoff frequency $f_{low} = 10^{-5}$ Hz and the initial observation frequency $f_{obs}=4.15\times10^{-5}(M_c/10^{6}M_\odot)^{-5/8}(T_{obs}/1 yr)^{-3/8}$ Hz.

Following the error strategy proposed in the literature \citep{Sathyaprakash2010,Zhao2011}, the total uncertainty of $D_{L}$ includes the contribution from the measurement itself $\sigma^{inst}_{D_L}$ and an additional uncertainty $\sigma^{lens}_{D_L}$ due to weak lensing:
\begin{align}
\sigma_{D_{L}}^{GW}=&~~\sqrt{\left(\sigma_{D_{L}}^{inst}\right)^{2}+\left(\sigma_{D_{L}}^{lens}\right)^{2}} \\
&~~=\sqrt{\left(\frac{2D_{L}}{\rho}\right)^{2}+\left(0.05zD_{L}\right)^{2}}.
\end{align}
The luminosity distance and total error of GW data can be obtained by the above process. Then we can test the spatial flatness of the universe in a cosmological-model-independent way using the GW data.

\section{Methodology and constraints on cosmic curvature} \label{sec.result}

\subsection{The GP method}

The GP method \citep{Seikel2012,Wu2020,Liu2019,Liao2019b,Liao2020} is a non-parametric smoothing method for reconstructing functions, which
does not need to consider the influence of other cosmological background models when placing limits on cosmic curvature. Especially, the distribution of $D_L(z)$ with redshift is obtained by GP method directly from $H(z)$ observations. In this work, we reconstruct the function $E(z)$ from the $H(z)$ data by using the GP method based on Gapp code, which has been
widely used in cosmological studies \citep{Wu2020,Liu2019}, and the $H(z)$ data are taken from Table 1 of \citet{Wei2017}. The advantage of this method is that we can get a model-independent function $E(z)$ without any prior assumption on the cosmological models \citep{Seikel2012,Cao2011,Cao2015,Cao2013}. Through the data analysis, the $H(z)$ can be obtained \textbf{in two ways: one by calculating the passively evolving galaxies's ages, and the other is based on the detection of radial BAO features}. We show the data in Table.~\ref{hzdata} \citep{Jimenez2003,Simon2005,Stern2010,Chuang2012,Moresco2012,Zhang2014,Moresco2015,Moresco2016,Gaztanaga2009,Blake2012,Samushia2013}.
Because the proper distance $d_P$ depends on the $E(z)$ function. We can use the $H(z)$ measurements to reconstruct the $E(z)$ function, and then derive $d_P$ from the function $E(z)$. The details of the steps to reconstruct the function are as follows: firstly, we normalize the $H(z)$ function and their $2\sigma$ errors based on the $H(z)$ data points by GP method. Secondly, the reconstructed function $E(z)$ can be obtained according to the $E(z)=H(z)/H_{0}$ relation where $H_0=69.6\pm0.7km/s/Mpc$ \citep{Bennett2014}. Thirdly, the proper distance $d_P$ with their $2\sigma$ errors can be obtained through the reconstructed function $E(z)$ using this formula $d_{P}(z)=\frac{c}{H_{0}}\int^{z}_{0}\frac{dz'}{E(z')}$. As is shown in Fig.~\ref{Ezdp_real}, the observed (blue points) and reconstructed (red solid lines) $H(z)$ are well consistent with those determined from the best-fit flat $\Lambda$CDM model (blue solid lines). The reconstruction of $d_P(z)$ function based on the $H(z)$ sample is also presented in Fig.~\ref{Ezdp_real}.

\subsection{Observational constraints on $\Omega_{k}$}

\begin{figure*}[htbp]
	\begin{center}
		\footnotesize
		\begin{tabular}{ccc}
			\includegraphics[width=0.5\textwidth]{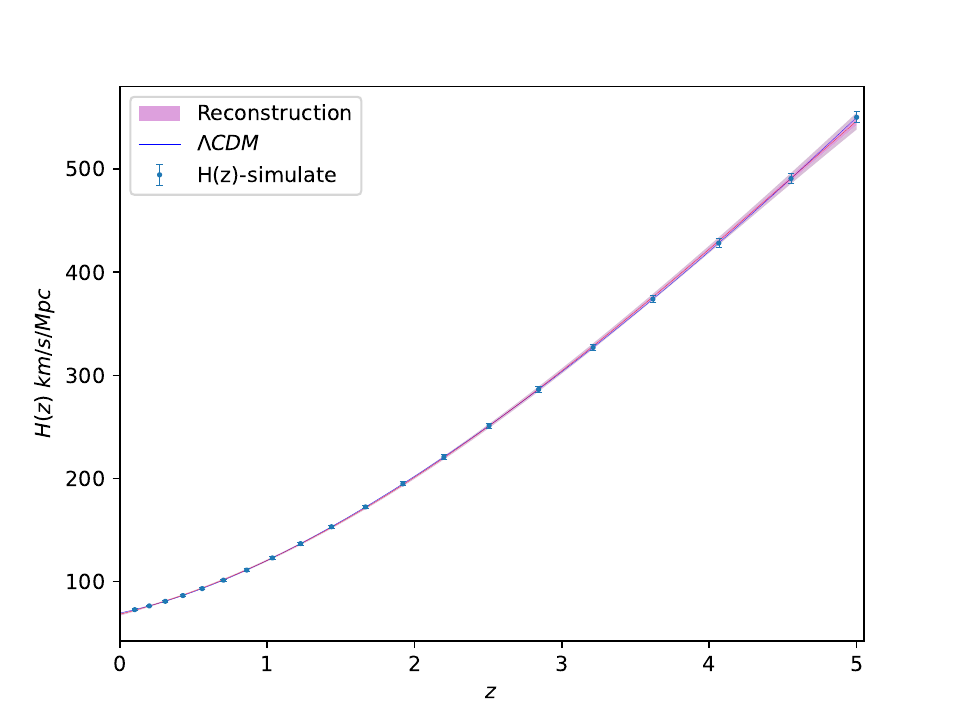} &
			\includegraphics[width=0.5\textwidth]{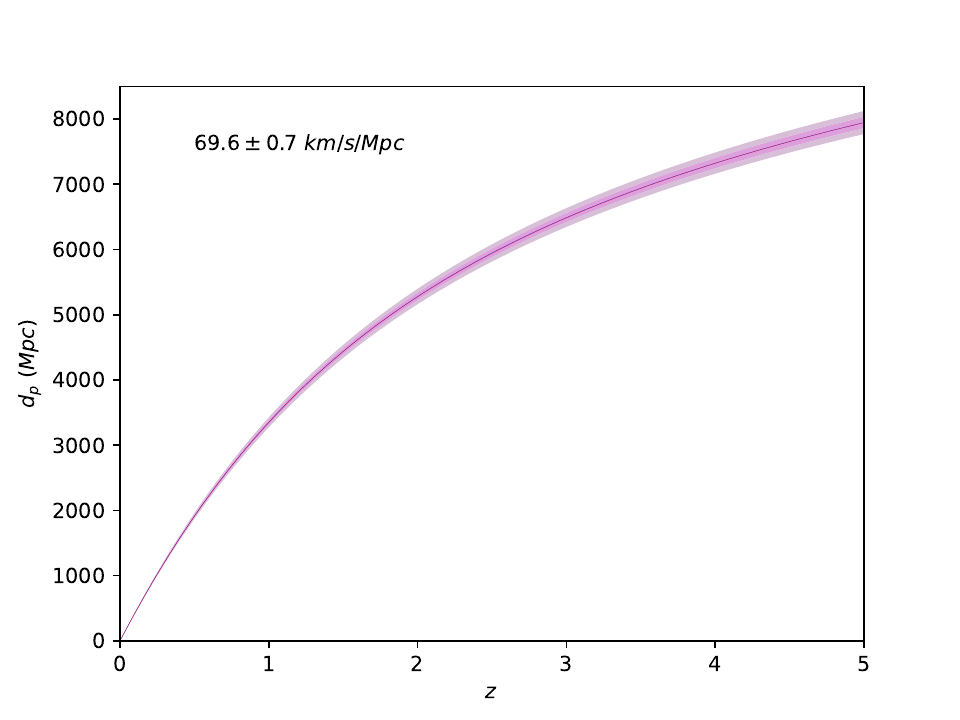} &\\
			(a) & (b)\\
		\end{tabular}
	\end{center}
	\caption{The reconstructed $H(z)$ and $d_p(z)$ function by GP method using the $H(z)$ simulated data with $H_0=69.6\pm0.7$ km/s/Mpc. The $H(z)$ simulated data and the $\Lambda$CDM model is also added for comparison. }
	\label{Ezdp_simulate}
\end{figure*}

\begin{figure}
	\includegraphics[width=0.5\textwidth]{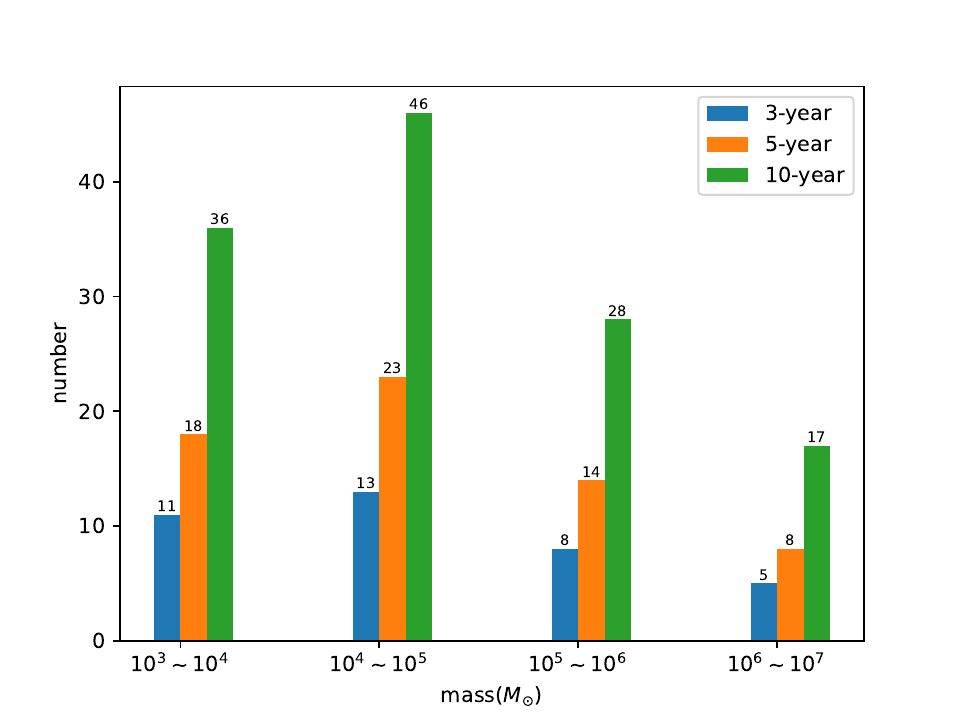}
	\caption{The specific number of SMBHBs that TQ could detect.}
	\label{number-of-TQ}
\end{figure}

\begin{figure*}[htbp]
	\begin{center}
		\footnotesize
		\begin{tabular}{ccc}
			\includegraphics[width=0.5\textwidth]{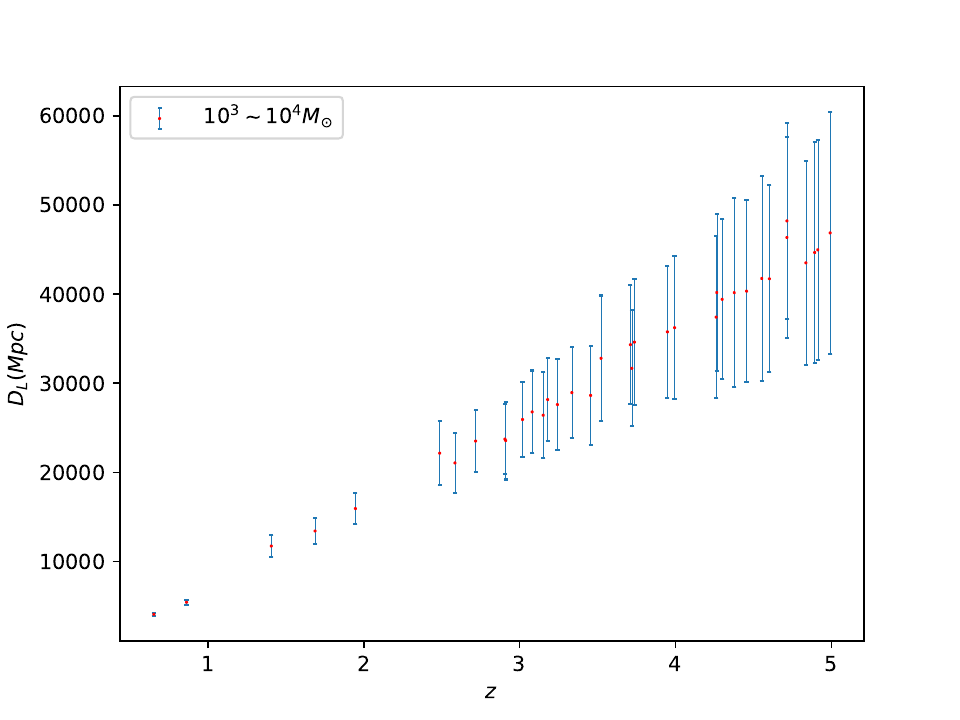} &
			\includegraphics[width=0.5\textwidth]{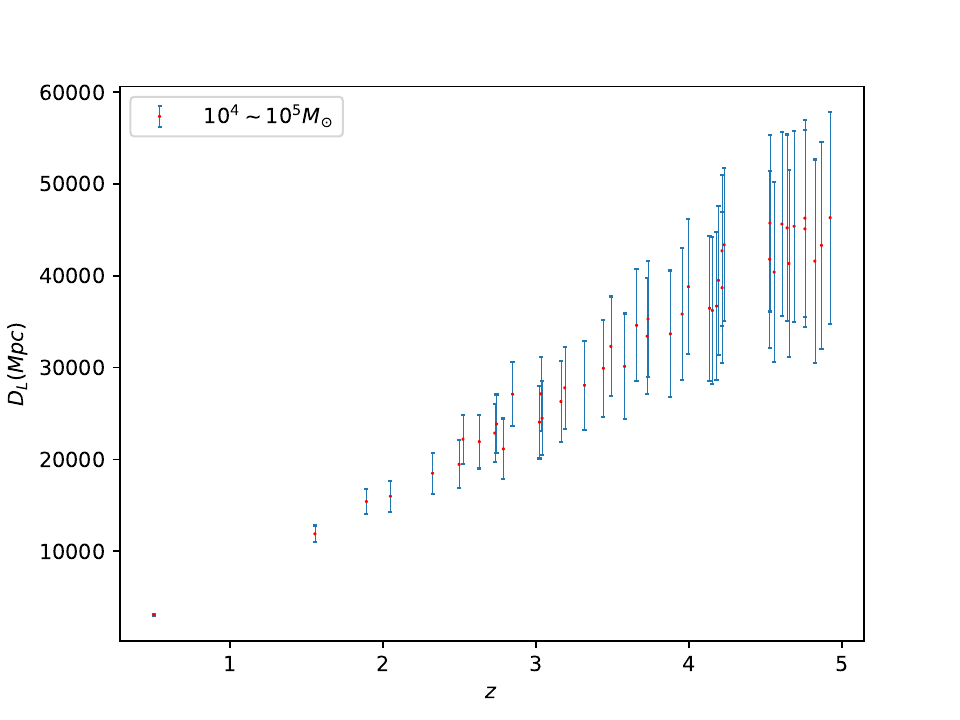} &\\
			(a) & (b)\\
			\includegraphics[width=0.5\textwidth]{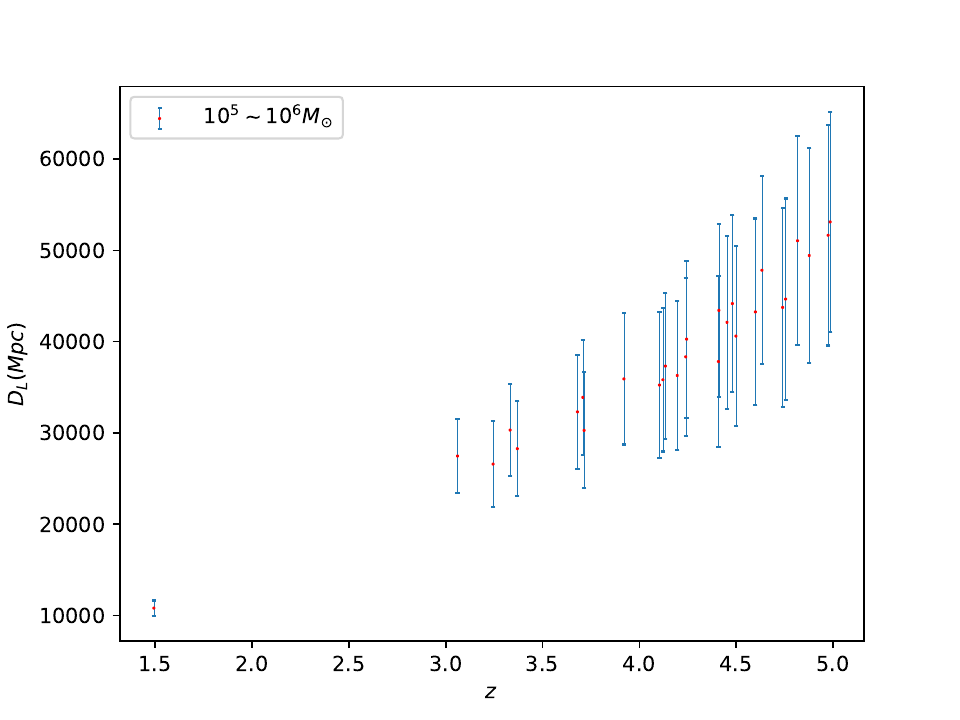} &
			\includegraphics[width=0.5\textwidth]{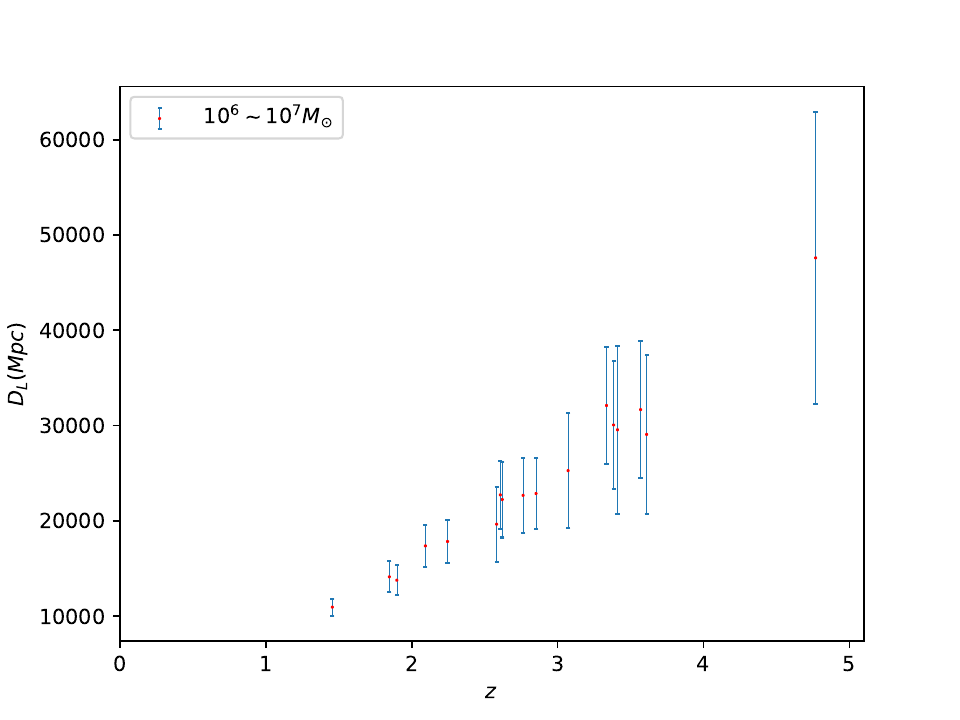} &\\
			(c) & (d)\\
		\end{tabular}
	\end{center}
	\caption{The simulated gravitational wave data ($D_{L}-z$ relation) derived from different sub-samples of SMBHBs (with total masses of $TQ10^{3}\sim 10^{7}M_\odot$), based on 10-year observation of TQ.}
	\label{z-distribution}
\end{figure*}

\begin{table*}[htbp]
	\begin{center}
		\footnotesize
		\caption{The best-fit spatial curvature ($\Omega_{k}$) and the corresponding $1\sigma$ uncertainties derived from different sub-samples of SMBHBs, based on  $H(z)$ simulated data and 3-, 5-, and 10-year observations of TQ.}
		\begin{tabular}{cccc}
			\hline
			$H(z)_{simulate}$+TQ ($M_\odot$)               &~~$\Omega_k$(3-year)~~   &~~$\Omega_k$(5-year)~~   &~~$\Omega_k$(10-year)~~\\
			\hline
			$10^{3}\sim 10^{4}$     &$0.02\pm0.23(1\sigma)$   &$0.00\pm0.16(1\sigma)$ &$-0.02\pm0.13(1\sigma)$ \\
			
			$10^{4}\sim 10^{5}$     &$-0.04\pm0.19(1\sigma)$    &$0.04\pm0.14(1\sigma)$  &$-0.014\pm0.097(1\sigma)$  \\
			
			$10^{5}\sim 10^{6}$    &$-0.03\pm0.25(1\sigma)$  &$-0.04\pm0.17(1\sigma)$  &$-0.01\pm0.12(1\sigma)$ \\
			
			$10^{6}\sim 10^{7}$    &$-0.04\pm0.4(1\sigma)$  &$-0.05\pm0.31(1\sigma)$ &$-0.06\pm0.2(1\sigma)$  \\
			
			$10^{3}\sim 10^{7}$    &$0.00\pm0.12(1\sigma)$   &$0.00\pm0.086(1\sigma)$ &$-0.002\pm0.061(1\sigma)$ \\
			\hline
		\end{tabular}
		\label{TianQin}
	\end{center}
\end{table*}

We adopt the $\chi^2$ minimum fitting method and Markov chain Monte Carlo (MCMC) technique to constrain the cosmic curvature $\Omega_{k}$, based on the following expression
\begin{equation}\label{HSFR}
\frac{D_{L}(z)}{(1+z)} = \left\lbrace \begin{array}{lll} \frac{c}{H_{0}}\frac{1}{\sqrt{|\Omega_{k}|}}\sinh\left[\sqrt{|\Omega_{k}|}d_{P}(z)\frac{H_{0}}{c}\right]~~{\rm for}~~\Omega_{k}>0\\
                                         d_{P}(z)~~~~~~~~~~~~~~~~~~~~~~~~~~~~~~~~~~~~{\rm for}~~\Omega_{k}=0\;, \\
                                         \frac{c}{H_{0}}\frac{1}{\sqrt{|\Omega_{k}|}}\sin\left[\sqrt{|\Omega_{k}|}d_{P}(z)\frac{H_{0}}{c}\right]~~~~{\rm for}~~\Omega_{k}<0\\
\end{array} \right.
\end{equation}
where $\Omega_k$ is the spatial curvature, $d_p$ is the proper distance that can be reconstructed by the GP method using $H(z)$ data. The corresponding error can be expressed as
\begin{equation}
\sigma_{D_{L}} = \left\lbrace \begin{array}{lll} (1+z)\cosh\left[\sqrt{|\Omega_{k}|}d_{P}(z)\frac{H_{0}}{c}\right]\sigma_{d_{P}}~~{\rm for}~~\Omega_{k}>0\\
                                         (1+z)\sigma_{d_{P}}~~~~~~~~~~~~~~~~~~~~~~~~~~~~~~~~{\rm for}~~\Omega_{k}=0\;, \\
                                         (1+z)\cos\left[\sqrt{|\Omega_{k}|}d_{P}(z)\frac{H_{0}}{c}\right]\sigma_{d_{P}}~~~~{\rm for}~~\Omega_{k}<0\\
\end{array} \right.
\end{equation}
The $\chi^2$ can be written as
\begin{equation}\label{dM}
    \chi^2=\sum_{i=1}^{N}\frac{(D_{L}^{obs}-D_{L}^{th})^2}{(\sigma_{D_L})^2+(\sigma_{D_L}^{GW})^2},
\end{equation}
where $N$ represents the total number of simulated GW data, $D_{L}^{obs}$ represents the observed luminosity distance, and the $D_{L}^{th}$ is the theoretical counterpart which can be obtained from Eq.~(\ref{HSFR}).

\textbf{In Fig.~\ref{TianQinP} (a), we show the probability distribution of the spatial curvature based on the 10-year observation of TQ ($10^{3}\sim 10^{7}M_\odot$) and $H(z)$ data. The best-fit curvature ($\Omega_k$) and its corresponding $1\sigma$ uncertainty are determined as $\Omega_k = 0.00\pm 0.19$. Due to the limited amount of TQ data simulated by popIII model within the mass range of $10^{3}\sim 10^{7}M_\odot$ and redshift range of $0\sim2.5$ (only 24 data points in 10 years), this result cannot fully demonstrate the constraining effect of TQ GW data on curvature. Therefore, we simulated $H(z)$ data to extend its upper limit of redshift, thereby expanding the dataset corresponding to TQ. We all acknowledge that the measurement of the BAO scale in the line-of-sight direction enables the determination of the Hubble parameter $H(z)$ at various redshifts. Consequently, \cite{Seo2007} employed even binning in $ln(1+z)$ to predict an all-sky survey within a redshift range of $0\sim 5$. Building upon the work of \cite{Seo2007}, \cite{Weinberg2013} utilizes their fast approximation method for the full Fisher matrix calculation to make predictions regarding $H(z)$. This method takes into consideration the idealized treatment of acoustic oscillations, non-linear structure formation, and redshift-space distortions. By incorporating survey redshift, number density, and volume, the method enables precise forecasts of $H(z)$ with an uncertainty of approximately $1\%$. Consequently, \cite{Yu2016} performed simulations of $H(z)$ data within the $0.1\sim 5$ redshift range, based on this approach, and constrained the cosmic curvature using a model-independent methodology. We refer to their work, and the specific process of $H(z)$ simulation in this paper is as follows: we adopted the flat $\Lambda$CDM model with parameter values $H_0 = 69.6~km/s/Mpc$ and $\Omega_m = 0.286$. A total of 20 data points were generated, evenly distributed in the $ln(1+z)$ space, covering a redshift range of $0.1\geq z \leq 5$ \citep{Yu2016},and the uncertainty of these relevant data is $1\%$ \citep{Weinberg2013}. Fig.~\ref{Ezdp_simulate} (a) shows the reconstruction results (solid red lines) of the simulated $H(z)$ data (blue points) using the GP method, which are consistent with the flat $\Lambda$CDM model (blue solid line). Fig.~\ref{Ezdp_simulate} (b) illustrates the reconstructed $d_p(z)$ function based on the simulated $H(z)$ data. At the same time, the redshift z range of $0\sim 5$ was selected for re-simulate the TQ GW data, following the same procedure as described in Chapter \ref{sec.data}. Additionally, Fig.\ref{number-of-TQ} presents the specific number distribution for each subsample simulated based on the popIII model, while Fig.\ref{z-distribution} displays the distribution of redshift ($z$) and luminosity distance ($D_L$) for the 10-year data.}

\begin{figure*}[htbp]
\begin{center}
\footnotesize
\begin{tabular}{ccc}
\includegraphics[width=0.5\textwidth]{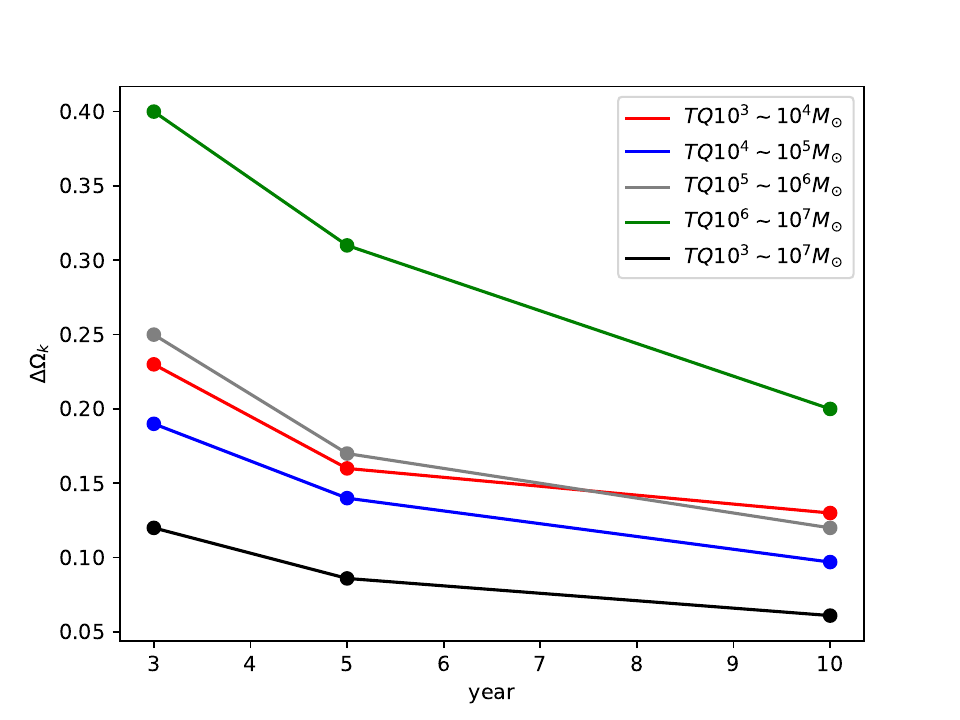} &
\includegraphics[width=0.5\textwidth]{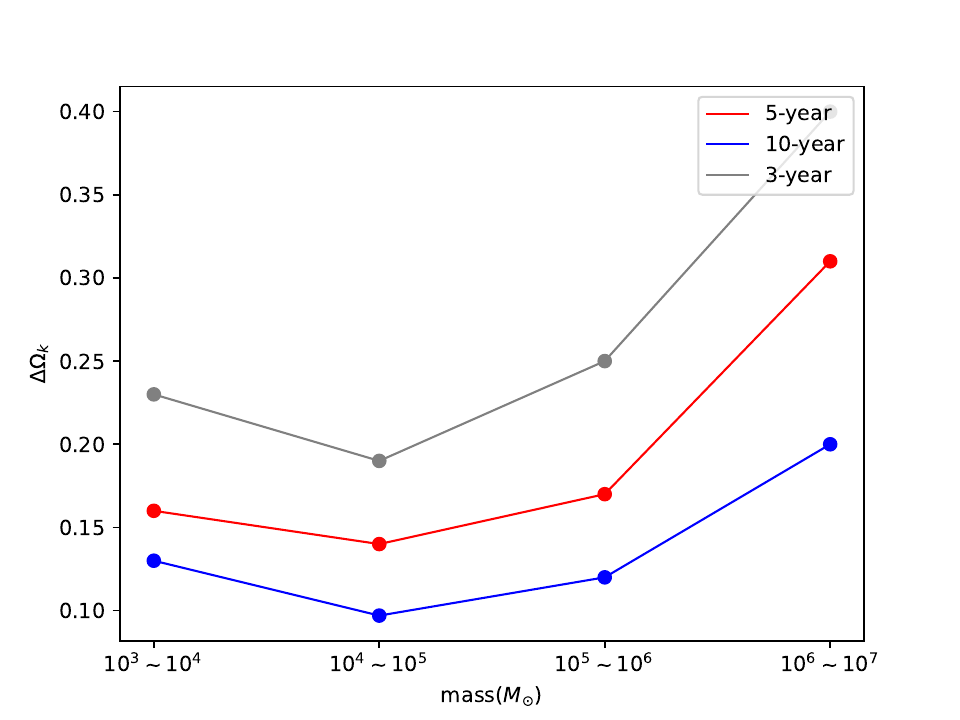} &\\
(a) & (b)\\
\end{tabular}
\end{center}
\caption{The variation of $\Omega_k$ precision with different observation time (left panel) and different total masses of SMBHB (right panel).}
\label{f3}
\end{figure*}

\begin{figure*}[htbp]
	\begin{center}
		\footnotesize
		\begin{tabular}{ccc}
			\includegraphics[width=0.5\textwidth]{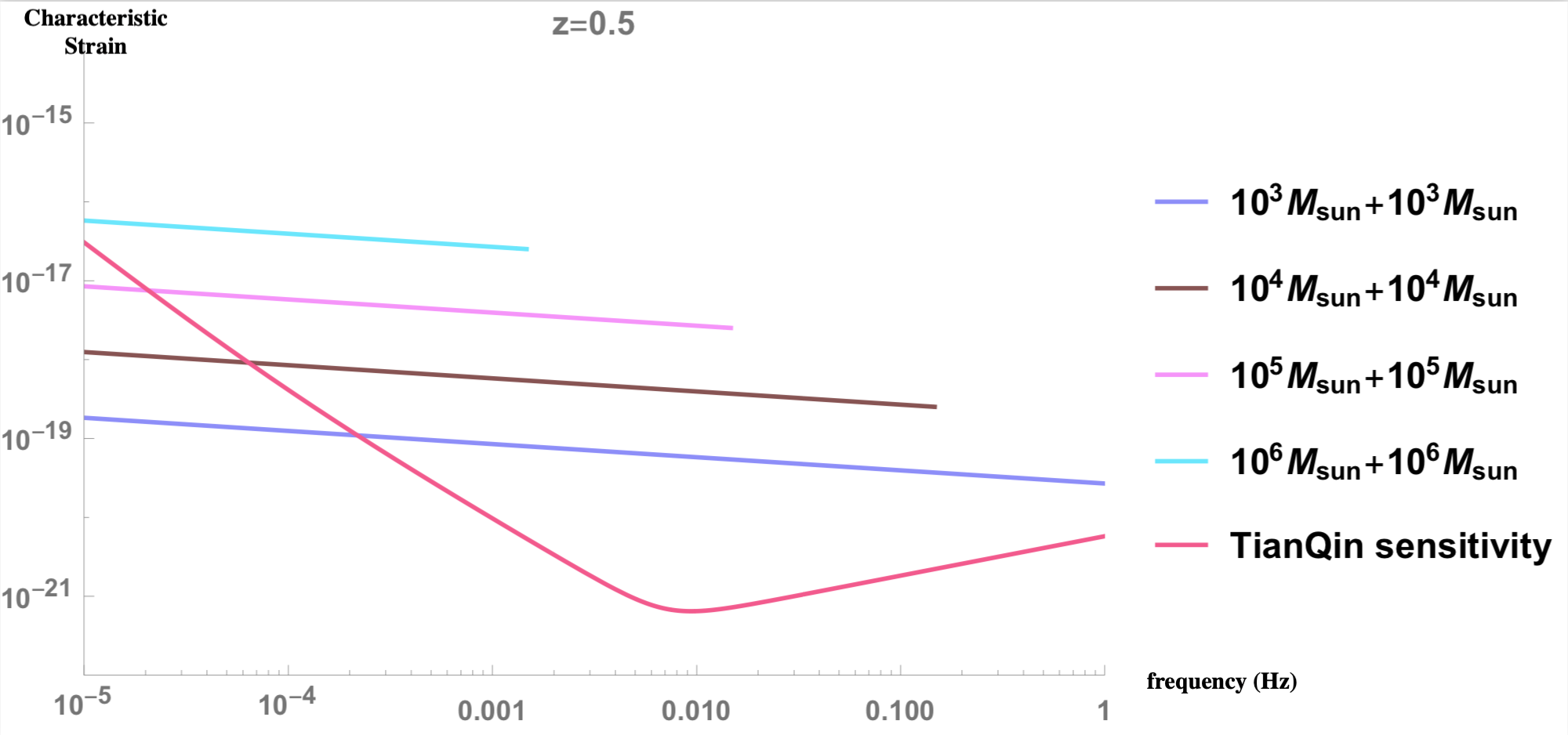} &
			\includegraphics[width=0.5\textwidth]{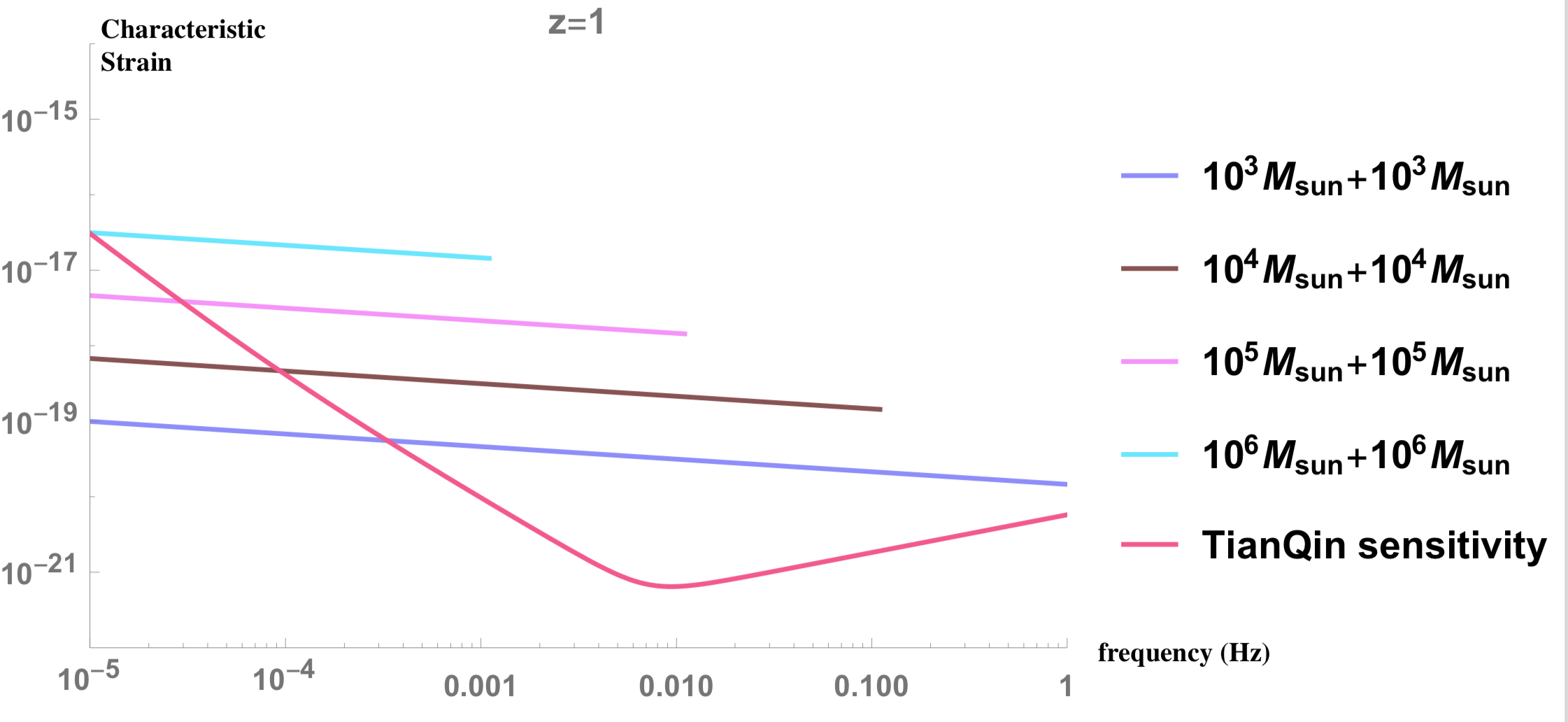} &\\
			(a) & (b)\\
			\includegraphics[width=0.5\textwidth]{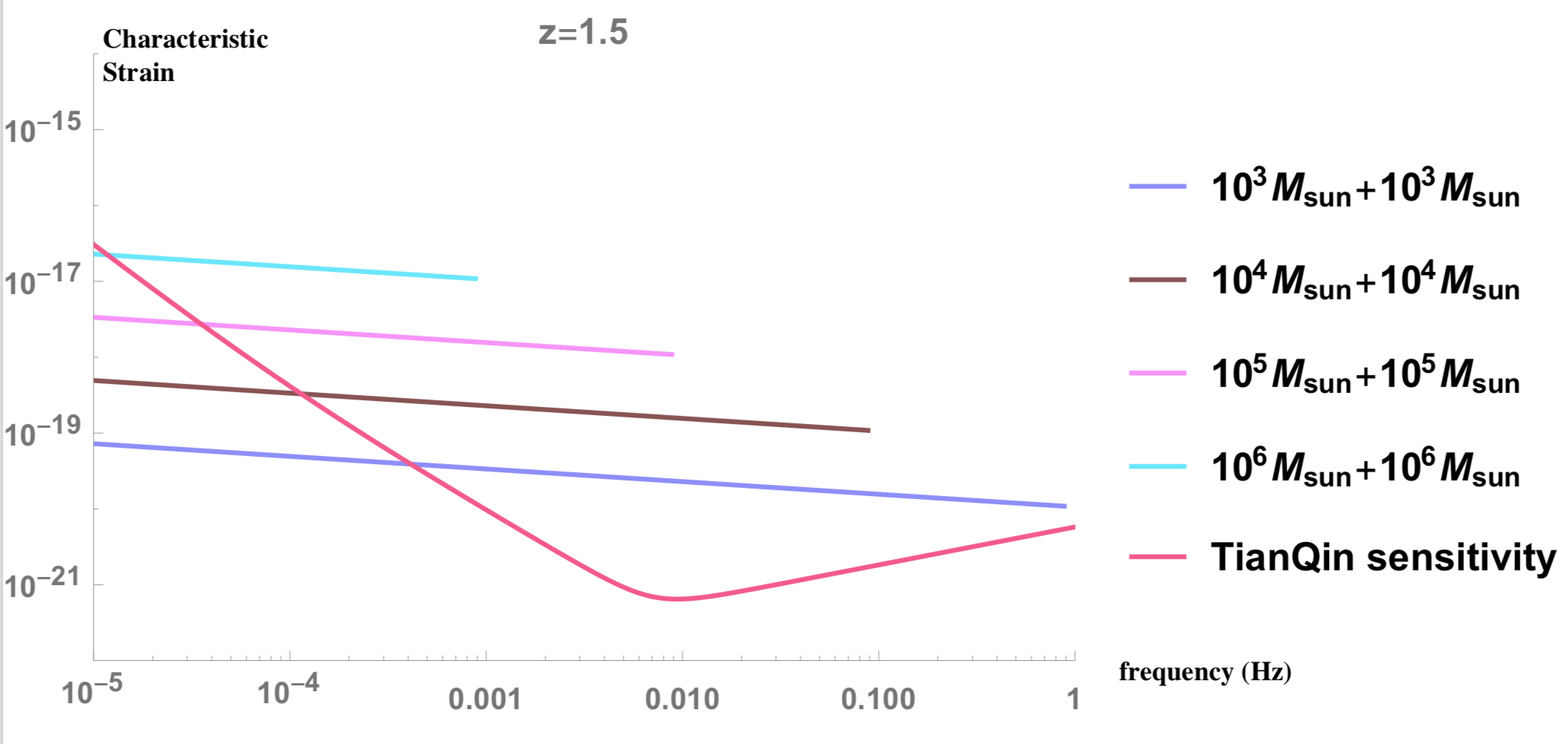} &
			\includegraphics[width=0.5\textwidth]{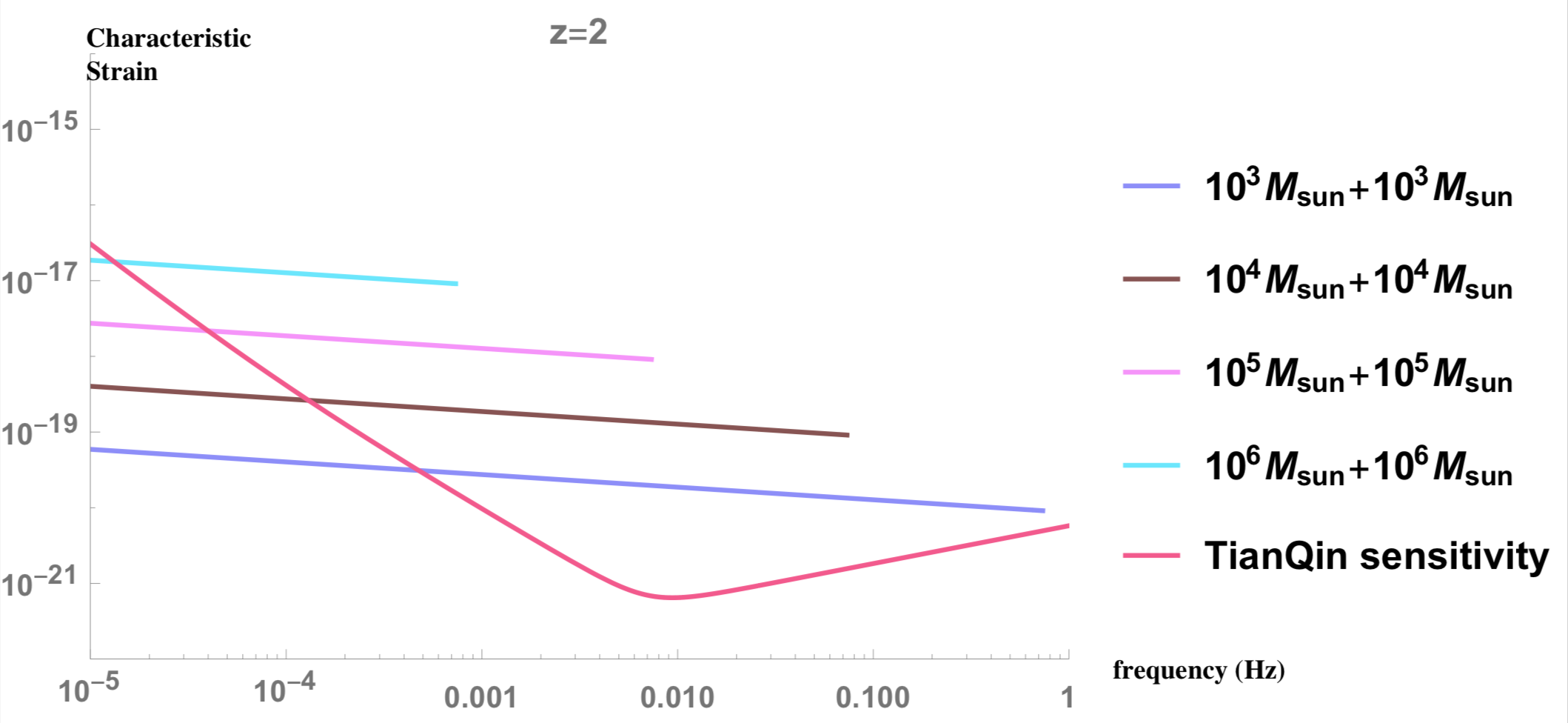} &\\
			(c) & (d)\\
		\end{tabular}
	\end{center}
	\caption{The amplitude evolution for SMBHBs with different mass ranges at redshift $z=0.5, 1, 1.5, 2$. The sensitivity curve of TQ is also added for comparison.}
	\label{sensitivity}
\end{figure*}

%In the simulation process, we adopted the flat $\Lambda$CDM model with parameter values $H_0 = 69.6~km/s/Mpc$ and $\Omega_m = 0.286$. A total of 20 data points were generated, evenly distributed in the $log(1+z)$ space, covering a redshift range of $0.1\geq z \leq 5$ \citep{Yu2016}. Furthermore, after analyzing many observational methods individually, \cite{Weinberg2013} developed a program to balance multiple techniques at an appropriate level of precision, for which the Fisher matrix was constructed to make predictions about cosmological parameters. The prediction results of $H(z)$ by using the fiducial BAO program show that the uncertainty of $H(z)$ is approximately $1\%$. Therefore, this level of uncertainty is incorporated into the simulation data. 
\textbf{Fig.~\ref{TianQinP} (b), (c) and (d) show the probability distribution of the spatial curvature based on the simulated $H(z)$ data and the 3-, 5- and 10-year observations of TQ, respectively.} Table.~\ref{TianQin} presents the best-fit curvature ($\Omega_k$) and the corresponding $1\sigma$ uncertainties obtained from different sub-samples. From the table, we can see that when the GW source total mass is different, the results obtained by using 10-year data are $-0.02\pm 0.13 (10^{3}\sim 10^{4}M_\odot)$, $-0.014\pm 0.097 (10^{4}\sim 10^{5}M_\odot)$, $-0.01\pm 0.12(10^{5}\sim 10^{6}M_\odot)$ and $-0.06\pm 0.2(10^{6}\sim 10^{7}M_\odot)$, respectively. In addition, we combine the $10^3M_\odot\sim10^7M_\odot$ data and the constraint on the parameter is $-0.002\pm 0.061$. These results all include $\Omega_k=0$ within the $1\sigma$ uncertainties, which means our universe is spatially flat. $\Omega_k=0$ is also included in the $1\sigma$ uncertainties for 3-, and 5-year's results. Moreover, through the comparison of analysis results in Fig.~\ref{f3} (a), we find that for GW sources with the same mass, the precision of cosmic curvature constraint would increase with the observation time. In Fig.~\ref{f3} (b), we also present the variation of cosmic curvature precision with the total mass of GW sources. Our results demonstrate an obvious $\Omega_k$ improvement when the total mass of GW source is $10^{4}\sim 10^{5}M_\odot$. There are two possible explanations for such interesting tendency: (I) GW sources with masses of $10^{4}\sim 10^{5}M_\odot$ dominate the full sample (as can be seen in Fig.~\ref{number-of-TQ}), which leads to the most stringent limits on the cosmic curvature; (II) The differences of SNR in SMBHB sub-samples covering different total masses could also potentially affect the constraints on $\Omega_k$. In order to investigate this reason, a plot of amplitudes of GW waveforms for SMBHBs with different mass ranges is shown in Fig.~\ref{sensitivity}, along with the TianQin sensitivity curve at redshift $z=0.5, 1, 1.5, 2$. Our findings suggest that the SMBHB sub-sample with total mass of $10^{4}\sim 10^{5}M_\odot$ posses very high sensitivity of detection, in the framework of the TianQin detectors.

In addition, in order to more intuitively reflect the constraint effect of TQ data on cosmic curvature, we compared the results obtained from other observation data constrained by the same method\citep{Zhang2022,Cao2022,Wei2018,Wei2020,Wei2020a}, the results are shown in Table.~\ref{comparison}. Firstly, we can see that the results given by the second-generation GW detector DECIGO and the third-generation GW detector ET are $\Omega_k = -0.05\pm0.12$ and $\Omega_k = 0.035\pm0.039$, respectively. Compared with DECIGO and ET's constraint results, the precision of our result ($\Omega_k=-0.002\pm 0.061$) is improved by $49.2\%$ and decreased by $56.4\%$, respectively. This indicates that the TQ is expected to provide a powerful and competitive probe of the spatial geometry of the universe, compared to future space-based detectors such as DECIGO. However, combining with the DECIGO and LISS strong lenses data, the result is $\Omega_k = 0.0001\pm0.012$, and the precision of our result is reduced by a factor of 4.1 compared to that. Besides, we also show the results given by Quasar ($\Omega_k = -0.918\pm0.429$) and Lensing+Quasar+SNe ($\Omega_k = 0.05^{+0.16}_{-0.14}$). We can see that the constraint precision of TQ GW data is $85.8\%$ and $59.3.3\%$ higher than theirs, respectively. This means that the GW data has a better constraint effect on cosmic curvature than electromagnetic wave data such as Quasar, Lenses, and SNe under this method. Finally, the precision of our results is reduced by a factor of 2.4 compared to Planck 2018 results ($\triangle\Omega_k=0.018$) \citep{Aghanim2020}.

\begin{table}
\begin{center}
\caption{The results of $\Omega_{k}$ were obtained by using the GP method.}
\begin{tabular}{lll}
\hline
Data                   &~~$\Omega_k$~~   &~~Source~~\\
\hline
TQ       &$-0.002\pm0.061$    & This Work\\
DECIGO    &$-0.05\pm0.12$     &\cite{Zhang2022}\\
DECIGO+LSST   &$0.0001\pm0.012$   &\cite{Cao2022}\\
ET       &$0.035\pm0.039$    &\cite{Wei2018}\\
Quasar      &$-0.918\pm0.429$  & \cite{Wei2020}\\
Lensing+Quasar+SNe      &$0.09\pm0.25$ & \cite{Wei2020a}  \\
\hline
\end{tabular}
\label{comparison}
\end{center}
\end{table}
%
%______________________________________________________________

\section{Conclusions}\label{sec.conclusion}

   \begin{enumerate}

      \item In this work, we investigate the spatial curvature of the universe in a GP way, based on the GW observations from the future space-based GW detector TQ. For the distance measurement unaffected by the cosmic opacity, we consider the simulated data of GWs from SMBHBs, which can be considered as standard sirens covering different redshift ranges. For the non-parametric reconstruction of distance and redshift relation, we use the $H(z)$ data, based on the non-parametric smoothing method called Gaussian Processes, and the reconstructed function is model-independent. Then we obtain the comoving distance $d_P(z)$ by directly solving the integral of this reconstructed function $E(z)$. Furthermore, with the spatial curvature $\Omega_k$ taken into consideration, one could transform the comoving distance into the $D_L$. The 1D distribution of cosmic curvature are shown in Fig.~\ref{TianQinP} (a), and the result is $\Omega_k = 0.00\pm 0.19$. \textbf{However, by analyzing the data, we found that the obtained result failed to demonstrate the constraint effect of TQ GW data on curvature. Therefore, we increased the upper limit of redshift by simulating $H(z)$ data to obtain additional TQ GW data. The final results are presented in Fig.\ref{TianQinP} (b), (c), and (d), along with the corresponding numerical estimations summarized in Table.\ref{TianQin}.}
      
      \item Our results show that $\Omega_k = -0.002\pm 0.061(1\sigma)$, by combining $H(z)$ simulated data with simulated GW events of TQ, and suggest that our universe is flat. In addition, according to the results in Table.~\ref{TianQin}, it is found that the precision of $\Omega_k$ constraint increases with the increase of observation duration, and the total mass of SMBHB does influence the estimation of cosmic curvature, implied by the analysis performed on different subsamples of gravitational wave data. Finally, compared to the previous model-independent constraints on the spatial curvature by GW data (DECIGO) \citep{Zhang2022,Wei2018}, TQ is expected to provide a powerful and competitive probe of the spatial geometry of the universe. The results are also available for the space-based GW detector Taiji, which will be sensitive in observing GWs from SMBHBs around 1 mHz.

      \item Through the analysis of the GW data from SMBHB observed by future GW detectors, our studies provide a possible approach to testing the spatial properties of the universe. Focusing on the strong degeneracy between the cosmic curvature and the Hubble constant \citep{Collett2019}, our results will be very helpful for us to investigate other fundamental issues in cosmology (such as the tension between $\Lambda$CDM model and astrophysical observations) \citep{Ding2015,Zheng2016,Qi2018}. We plan to explore these possibilities in the future works.

   \end{enumerate}

\begin{acknowledgements}
We appreciate Pro. Zong-Hong Zhu for useful discussions. This work was supported by National Natural Science Foundation of China (Grant Nos. 12105032, 11873001, 12047564, 12075041, 12147102), the National Key R\&D Program of China No. 2017YFA0402600; the National Natural Science Foundation of China under Grants No.11503001, 11690023, 11373014, 11873001 and 11633001; the Strategic Priority Research Program of the Chinese Academy of Sciences, Grant No. XDB23000000; the Interdiscipline Research Funds of Beijing Normal University, the Opening Project of Key Laboratory of Computational Astrophysics, National Astronomical Observatories, Chinese Academy of Sciences, and Chongqing Municipal Science and Technology Commission Fund(cstc2018jcyjAX0192). The Natural Science Foundation of Chongqing (Grant No. cstc2018jcyjAX0767). The Fundamental Research Funds for the Central Universities under Grant No. 2020CDJQY-Z003; the Fundamental Research Funds for the Central Universities of China No. 2021CDJQY-011, 2020CDJQYZ003; the Science Foundation of Chongqing under Grant No. D63012022005; the Natural Science Foundation of Chongqing No. cstc2021jcyj-msxmX0481;the Sichuan Youth Science and Technology Innovation Research Team (Grant No. 21CXTD0038)

\end{acknowledgements}


\bibliographystyle{aa} % style aa.bst
\bibliography{tianqin} % your references Yourfile.bib

\begin{thebibliography}{96}
\expandafter\ifx\csname natexlab\endcsname\relax\def\natexlab#1{#1}\fi

\bibitem[{Abbott {et~al.}(2017{\natexlab{a}})}]{Abbott2017a}
Abbott, B. . P.~. {et~al.} 2017{\natexlab{a}}, Astrophys. J. Lett., 851, L35

\bibitem[{Abbott {et~al.}(2016{\natexlab{a}})}]{Abbott2016b}
Abbott, B.~P. {et~al.} 2016{\natexlab{a}}, Phys. Rev. D, 94, 064035

\bibitem[{Abbott {et~al.}(2016{\natexlab{b}})}]{Abbott2016c}
Abbott, B.~P. {et~al.} 2016{\natexlab{b}}, Phys. Rev. Lett., 116, 241103

\bibitem[{Abbott {et~al.}(2016{\natexlab{c}})}]{Abbott2016a}
Abbott, B.~P. {et~al.} 2016{\natexlab{c}}, Astrophys. J. Lett., 826, L13

\bibitem[{Abbott {et~al.}(2016{\natexlab{d}})}]{Abbott2016}
Abbott, B.~P. {et~al.} 2016{\natexlab{d}}, Phys. Rev. Lett., 116, 061102

\bibitem[{Abbott {et~al.}(2017{\natexlab{b}})}]{Abbott2017}
Abbott, B.~P. {et~al.} 2017{\natexlab{b}}, Phys. Rev. Lett., 118, 221101,
  [Erratum: Phys.Rev.Lett. 121, 129901 (2018)]

\bibitem[{Abbott {et~al.}(2017{\natexlab{c}})}]{Abbott2017b}
Abbott, B.~P. {et~al.} 2017{\natexlab{c}}, Phys. Rev. Lett., 119, 141101

\bibitem[{Abbott {et~al.}(2018)}]{Abbott2018}
Abbott, B.~P. {et~al.} 2018, Phys. Rev. Lett., 121, 161101

\bibitem[{Abbott {et~al.}(2020)}]{Abbott2020}
Abbott, B.~P. {et~al.} 2020, Astrophys. J. Lett., 892, L3

\bibitem[{Aghanim {et~al.}(2020)}]{Aghanim2020}
Aghanim, N. {et~al.} 2020, Astron. Astrophys., 641, A6, [Erratum:
  Astron.Astrophys. 652, C4 (2021)]

\bibitem[{Bennett {et~al.}(2014)Bennett, Larson, Weiland, \&
  Hinshaw}]{Bennett2014}
Bennett, C.~L., Larson, D., Weiland, J.~L., \& Hinshaw, G. 2014, Astrophys. J.,
  794, 135

\bibitem[{Bernstein(2006)}]{Bernstein2006}
Bernstein, G. 2006, Astrophys. J., 637, 598

\bibitem[{Blake {et~al.}(2012)}]{Blake2012}
Blake, C. {et~al.} 2012, Mon. Not. Roy. Astron. Soc., 425, 405

\bibitem[{Boehm \& R\"as\"anen(2013)}]{Boehm2013}
Boehm, C. \& R\"as\"anen, S. 2013, JCAP, 09, 003

\bibitem[{Busca {et~al.}(2013)}]{Busca2013}
Busca, N.~G. {et~al.} 2013, Astron. Astrophys., 552, A96

\bibitem[{Cai {et~al.}(2016)Cai, Guo, \& Yang}]{Cai2016}
Cai, R.-G., Guo, Z.-K., \& Yang, T. 2016, Phys. Rev. D, 93, 043517

\bibitem[{Cai \& Yang(2017)}]{Cai2017}
Cai, R.-G. \& Yang, T. 2017, Phys. Rev. D, 95, 044024

\bibitem[{Cao {et~al.}(2015)Cao, Chen, Zhang, \& Ma}]{Cao2015}
Cao, S., Chen, Y., Zhang, J., \& Ma, Y. 2015, Int. J. Theor. Phys., 54, 1492

\bibitem[{Cao {et~al.}(2017)Cao, Li, Biesiada, Xu, Cai, \& Zhu}]{Cao2017}
Cao, S., Li, X., Biesiada, M., {et~al.} 2017, Astrophys. J., 835, 92

\bibitem[{Cao \& Liang(2013)}]{Cao2013}
Cao, S. \& Liang, N. 2013, International Journal of Modern Physics D, 22,
  1350082

\bibitem[{Cao {et~al.}(2011)Cao, Liang, \& Zhu}]{Cao2011}
Cao, S., Liang, N., \& Zhu, Z.-H. 2011, Mon. Not. Roy. Astron. Soc., 416, 1099

\bibitem[{Cao {et~al.}(2022)Cao, Liu, Biesiada, Liu, Guo, \& Zhu}]{Cao2022}
Cao, S., Liu, T., Biesiada, M., {et~al.} 2022, Astrophys. J., 926, 214

\bibitem[{Cao {et~al.}(2019)Cao, Qi, Cao, Biesiada, Li, Pan, \& Zhu}]{Cao2019}
Cao, S., Qi, J., Cao, Z., {et~al.} 2019, Sci. Rep., 9, 11608

\bibitem[{{Chuang} \& {Wang}(2012)}]{Chuang2012}
{Chuang}, C.-H. \& {Wang}, Y. 2012, \mnras, 426, 226

\bibitem[{Clarkson {et~al.}(2007)Clarkson, Cortes, \& Bassett}]{Clarkson2007}
Clarkson, C., Cortes, M., \& Bassett, B.~A. 2007, JCAP, 08, 011

\bibitem[{Collett {et~al.}(2019)Collett, Montanari, \& Rasanen}]{Collett2019}
Collett, T., Montanari, F., \& Rasanen, S. 2019, Phys. Rev. Lett., 123, 231101

\bibitem[{Delubac {et~al.}(2015)}]{Delubac2015}
Delubac, T. {et~al.} 2015, Astron. Astrophys., 574, A59

\bibitem[{Denissenya {et~al.}(2018)Denissenya, Linder, \&
  Shafieloo}]{Denissenya2018}
Denissenya, M., Linder, E.~V., \& Shafieloo, A. 2018, JCAP, 03, 041

\bibitem[{Di \& Gong(2018)}]{Di2018}
Di, H. \& Gong, Y. 2018, JCAP, 07, 007

\bibitem[{Ding {et~al.}(2015)Ding, Biesiada, Cao, Li, \& Zhu}]{Ding2015}
Ding, X., Biesiada, M., Cao, S., Li, Z., \& Zhu, Z.-H. 2015, Astrophys. J.
  Lett., 803, L22

\bibitem[{Enqvist(2008)}]{Enqvist2008}
Enqvist, K. 2008, Gen. Rel. Grav., 40, 451

\bibitem[{Feng {et~al.}(2019)Feng, Wang, Hu, Hu, \& Wang}]{Feng2019}
Feng, W.-F., Wang, H.-T., Hu, X.-C., Hu, Y.-M., \& Wang, Y. 2019, Phys. Rev. D,
  99, 123002

\bibitem[{Ferrer {et~al.}(2009)Ferrer, Multamaki, \& Rasanen}]{Ferrer2009}
Ferrer, F., Multamaki, T., \& Rasanen, S. 2009, JHEP, 04, 006

\bibitem[{Ferrer \& Rasanen(2006)}]{Ferrer2006}
Ferrer, F. \& Rasanen, S. 2006, JHEP, 02, 016

\bibitem[{Font-Ribera {et~al.}(2014)}]{FontRibera2014}
Font-Ribera, A. {et~al.} 2014, JCAP, 05, 027

\bibitem[{Gaztanaga {et~al.}(2009)Gaztanaga, Cabre, \& Hui}]{Gaztanaga2009}
Gaztanaga, E., Cabre, A., \& Hui, L. 2009, Mon. Not. Roy. Astron. Soc., 399,
  1663

\bibitem[{Gong {et~al.}(2021)Gong, Luo, \& Wang}]{Gong2021}
Gong, Y., Luo, J., \& Wang, B. 2021, Nature Astron., 5, 881

\bibitem[{Gong \& Wang(2007)}]{Gong2007}
Gong, Y.-G. \& Wang, A. 2007, Phys. Rev. D, 75, 043520

\bibitem[{He {et~al.}(2022)He, Pan, Shi, Li, Cao, \& Cheng}]{He2022}
He, Y., Pan, Y., Shi, D., {et~al.} 2022, Res. Astron. Astrophys., 22, 085016

\bibitem[{Hughes(2001)}]{Hughes2001}
Hughes, S.~A. 2001, Class. Quant. Grav., 18, 4067

\bibitem[{Ichikawa {et~al.}(2006)Ichikawa, Kawasaki, Sekiguchi, \&
  Takahashi}]{Ichikawa2006}
Ichikawa, K., Kawasaki, M., Sekiguchi, T., \& Takahashi, T. 2006, JCAP, 12, 005

\bibitem[{Jimenez {et~al.}(2003)Jimenez, Verde, Treu, \& Stern}]{Jimenez2003}
Jimenez, R., Verde, L., Treu, T., \& Stern, D. 2003, Astrophys. J., 593, 622

\bibitem[{Klein {et~al.}(2016)Klein, Barausse, Sesana, Petiteau, Berti, Babak,
  Gair, Aoudia, Hinder, Ohme, \& Wardell}]{Klein2016}
Klein, A., Barausse, E., Sesana, A., {et~al.} 2016, Phys. Rev. D, 93, 024003

\bibitem[{Lavinto {et~al.}(2013)Lavinto, R\"as\"anen, \& Szybka}]{Lavinto2013}
Lavinto, M., R\"as\"anen, S., \& Szybka, S.~J. 2013, JCAP, 12, 051

\bibitem[{Li {et~al.}(2008)Li, Baker, Fang, Stephenson, \& Chen}]{Li2008}
Li, F., Baker, Jr., R. M.~L., Fang, Z., Stephenson, G.~V., \& Chen, Z. 2008,
  Eur. Phys. J. C, 56, 407

\bibitem[{Li {et~al.}(2003)Li, Tang, \& Shi}]{Li2003}
Li, F.-Y., Tang, M.-X., \& Shi, D.-P. 2003, Phys. Rev. D, 67, 104008

\bibitem[{Li {et~al.}(2014)Li, Li, Zhang, \& Li}]{Li2014}
Li, Y.-L., Li, S.-Y., Zhang, T.-J., \& Li, T.-P. 2014, Astrophys. J. Lett.,
  789, L15

\bibitem[{Li {et~al.}(2016)Li, Wang, Liao, \& Zhu}]{Li2016}
Li, Z., Wang, G.-J., Liao, K., \& Zhu, Z.-H. 2016, Astrophys. J., 833, 240

\bibitem[{Liao(2019{\natexlab{a}})}]{Liao2019}
Liao, K. 2019{\natexlab{a}}, Phys. Rev. D, 99, 083514

\bibitem[{Liao(2019{\natexlab{b}})}]{Liao2019a}
Liao, K. 2019{\natexlab{b}}, Astrophys. J., 885, 70

\bibitem[{Liao {et~al.}(2022)Liao, Biesiada, \& Zhu}]{Liao2022}
Liao, K., Biesiada, M., \& Zhu, Z.-H. 2022, Chin. Phys. Lett., 39, 119801

\bibitem[{Liao {et~al.}(2017{\natexlab{a}})Liao, Fan, Ding, Biesiada, \&
  Zhu}]{Liao2017a}
Liao, K., Fan, X.-L., Ding, X.-H., Biesiada, M., \& Zhu, Z.-H.
  2017{\natexlab{a}}, Nature Commun., 8, 1148, [Erratum: Nature Commun. 8, 2136
  (2017)]

\bibitem[{Liao {et~al.}(2017{\natexlab{b}})Liao, Li, Wang, \& Fan}]{Liao2017}
Liao, K., Li, Z., Wang, G.-J., \& Fan, X.-L. 2017{\natexlab{b}}, Astrophys. J.,
  839, 70

\bibitem[{Liao {et~al.}(2019)Liao, Shafieloo, Keeley, \& Linder}]{Liao2019b}
Liao, K., Shafieloo, A., Keeley, R.~E., \& Linder, E.~V. 2019, Astrophys. J.
  Lett., 886, L23

\bibitem[{Liao {et~al.}(2020)Liao, Shafieloo, Keeley, \& Linder}]{Liao2020}
Liao, K., Shafieloo, A., Keeley, R.~E., \& Linder, E.~V. 2020, Astrophys. J.
  Lett., 895, L29

\bibitem[{Liu {et~al.}(2019)Liu, Cao, Zhang, Geng, Liu, Ji, \& Zhu}]{Liu2019}
Liu, T., Cao, S., Zhang, J., {et~al.} 2019, Astrophys. J., 886, 94

\bibitem[{Luo {et~al.}(2016)Luo, Chen, Duan, Gong, Hu, Ji, Liu, Mei, Milyukov,
  Sazhin, Shao, Toth, Tu, Wang, Wang, Yeh, Zhan, Zhang, Zharov, \&
  Zhou}]{Luo2016}
Luo, J., Chen, L.-S., Duan, H.-Z., {et~al.} 2016, Classical and Quantum
  Gravity, 33, 035010

\bibitem[{{Moresco}(2015)}]{Moresco2015}
{Moresco}, M. 2015, \mnras, 450, L16

\bibitem[{Moresco {et~al.}(2016)Moresco, Pozzetti, Cimatti, Jimenez, Maraston,
  Verde, Thomas, Citro, Tojeiro, \& Wilkinson}]{Moresco2016}
Moresco, M., Pozzetti, L., Cimatti, A., {et~al.} 2016, Journal of Cosmology and
  Astroparticle Physics, 2016, 014

\bibitem[{Moresco {et~al.}(2012)Moresco, Verde, Pozzetti, Jimenez, \&
  Cimatti}]{Moresco2012}
Moresco, M., Verde, L., Pozzetti, L., Jimenez, R., \& Cimatti, A. 2012, JCAP,
  07, 053

\bibitem[{Mortsell \& Jonsson(2011)}]{Mortsell2011}
Mortsell, E. \& Jonsson, J. 2011 [\eprint[arXiv]{1102.4485}]

\bibitem[{Olmez {et~al.}(2010)Olmez, Mandic, \& Siemens}]{Olmez2010}
Olmez, S., Mandic, V., \& Siemens, X. 2010, Phys. Rev. D, 81, 104028

\bibitem[{Pan {et~al.}(2021)Pan, He, Qi, Li, Cao, Liu, \& Wang}]{Pan2021}
Pan, Y., He, Y., Qi, J., {et~al.} 2021, Astrophys. J., 911, 135

\bibitem[{Petiteau {et~al.}(2011)Petiteau, Babak, \& Sesana}]{Petiteau2011}
Petiteau, A., Babak, S., \& Sesana, A. 2011, Astrophys. J., 732, 82

\bibitem[{Qi {et~al.}(2019)Qi, Cao, Biesiada, Ding, Zhu, \& Zheng}]{Qi2019}
Qi, J., Cao, S., Biesiada, M., {et~al.} 2019, Phys. Rev. D, 100, 023530

\bibitem[{Qi {et~al.}(2018)Qi, Cao, Biesiada, Xu, Wu, Zhang, \& Zhu}]{Qi2018}
Qi, J.-Z., Cao, S., Biesiada, M., {et~al.} 2018, Res. Astron. Astrophys., 18,
  066

\bibitem[{Rasanen(2009)}]{Rasanen2009}
Rasanen, S. 2009, JCAP, 02, 011

\bibitem[{Redlich {et~al.}(2014)Redlich, Bolejko, Meyer, Lewis, \&
  Bartelmann}]{Redlich2014}
Redlich, M., Bolejko, K., Meyer, S., Lewis, G.~F., \& Bartelmann, M. 2014,
  Astron. Astrophys., 570, A63

\bibitem[{Samushia {et~al.}(2013)}]{Samushia2013}
Samushia, L. {et~al.} 2013, Mon. Not. Roy. Astron. Soc., 429, 1514

\bibitem[{Sapone {et~al.}(2014)Sapone, Majerotto, \& Nesseris}]{Sapone2014}
Sapone, D., Majerotto, E., \& Nesseris, S. 2014, Phys. Rev. D, 90, 023012

\bibitem[{Sathyaprakash {et~al.}(2010)Sathyaprakash, Schutz, \& Van
  Den~Broeck}]{Sathyaprakash2010}
Sathyaprakash, B.~S., Schutz, B.~F., \& Van Den~Broeck, C. 2010, Class. Quant.
  Grav., 27, 215006

\bibitem[{Schutz(1986)}]{Schutz1986}
Schutz, B.~F. 1986, Nature, 323, 310

\bibitem[{Seikel {et~al.}(2012)Seikel, Clarkson, \& Smith}]{Seikel2012}
Seikel, M., Clarkson, C., \& Smith, M. 2012, JCAP, 06, 036

\bibitem[{Seo \& Eisenstein(2007)}]{Seo2007}
Seo, H.-J. \& Eisenstein, D.~J. 2007, Astrophys. J., 665, 14

\bibitem[{Sesana(2016)}]{Sesana2016}
Sesana, A. 2016, Phys. Rev. Lett., 116, 231102

\bibitem[{Shafieloo \& Clarkson(2010)}]{Shafieloo2010}
Shafieloo, A. \& Clarkson, C. 2010, Phys. Rev. D, 81, 083537

\bibitem[{Simon {et~al.}(2005)Simon, Verde, \& Jimenez}]{Simon2005}
Simon, J., Verde, L., \& Jimenez, R. 2005, Phys. Rev. D, 71, 123001

\bibitem[{{Stern} {et~al.}(2010){Stern}, {Jimenez}, {Verde}, {Kamionkowski}, \&
  {Stanford}}]{Stern2010}
{Stern}, D., {Jimenez}, R., {Verde}, L., {Kamionkowski}, M., \& {Stanford},
  S.~A. 2010, \jcap, 2010, 008

\bibitem[{Virey {et~al.}(2008)Virey, Talon-Esmieu, Ealet, Taxil, \&
  Tilquin}]{Virey2008}
Virey, J.~M., Talon-Esmieu, D., Ealet, A., Taxil, P., \& Tilquin, A. 2008,
  JCAP, 12, 008

\bibitem[{Wei(2018)}]{Wei2018}
Wei, J.-J. 2018, Astrophys. J., 868, 29

\bibitem[{Wei \& Melia(2020{\natexlab{a}})}]{Wei2020a}
Wei, J.-J. \& Melia, F. 2020{\natexlab{a}}, Astrophys. J., 897, 127

\bibitem[{Wei \& Melia(2020{\natexlab{b}})}]{Wei2020}
Wei, J.-J. \& Melia, F. 2020{\natexlab{b}}, The Astrophysical Journal, 888, 99

\bibitem[{Wei \& Wu(2017)}]{Wei2017}
Wei, J.-J. \& Wu, X.-F. 2017, Astrophys. J., 838, 160

\bibitem[{Weinberg {et~al.}(2013)Weinberg, Mortonson, Eisenstein, Hirata,
  Riess, \& Rozo}]{Weinberg2013}
Weinberg, D.~H., Mortonson, M.~J., Eisenstein, D.~J., {et~al.} 2013, Phys.
  Rept., 530, 87

\bibitem[{Wright(2007)}]{Wright2007}
Wright, E.~L. 2007, Astrophys. J., 664, 633

\bibitem[{Wu {et~al.}(2020)Wu, Cao, Zhang, Liu, Liu, Geng, \& Lian}]{Wu2020}
Wu, Y., Cao, S., Zhang, J., {et~al.} 2020, The Astrophysical Journal, 888, 113

\bibitem[{Xu {et~al.}(2013)Xu, Cuesta, Padmanabhan, Eisenstein, \&
  McBride}]{Xu2013}
Xu, X., Cuesta, A.~J., Padmanabhan, N., Eisenstein, D.~J., \& McBride, C.~K.
  2013, Mon. Not. Roy. Astron. Soc., 431, 2834

\bibitem[{Yi {et~al.}(2022)Yi, Nelemans, Brinkerink, Kostrzewa-Rutkowska,
  Timmer, Stoppa, Rossi, \& Portegies~Zwart}]{Yi2022}
Yi, S.-X., Nelemans, G., Brinkerink, C., {et~al.} 2022, Astron. Astrophys.,
  663, A155

\bibitem[{Yu \& Wang(2016)}]{Yu2016}
Yu, H. \& Wang, F.~Y. 2016, Astrophys. J., 828, 85

\bibitem[{{Zhang} {et~al.}(2014){Zhang}, {Zhang}, {Yuan}, {Liu}, {Zhang}, \&
  {Sun}}]{Zhang2014}
{Zhang}, C., {Zhang}, H., {Yuan}, S., {et~al.} 2014, Research in Astronomy and
  Astrophysics, 14, 1221

\bibitem[{Zhang {et~al.}(2023)Zhang, Diao, Pan, Cheng, \& Li}]{Zhang2023}
Zhang, J.-W., Diao, J., Pan, Y., Cheng, M.-Y., \& Li, J. 2023, Chin. Phys. C,
  47, 035103

\bibitem[{Zhang {et~al.}(2022)Zhang, Cao, Liu, Liu, Liu, \& Zheng}]{Zhang2022}
Zhang, Y., Cao, S., Liu, X., {et~al.} 2022, Astrophys. J., 931, 119

\bibitem[{Zhao {et~al.}(2007)Zhao, Xia, Li, Tao, Virey, Zhu, \&
  Zhang}]{Zhao2007}
Zhao, G.-B., Xia, J.-Q., Li, H., {et~al.} 2007, Phys. Lett. B, 648, 8

\bibitem[{Zhao {et~al.}(2011)Zhao, Van Den~Broeck, Baskaran, \& Li}]{Zhao2011}
Zhao, W., Van Den~Broeck, C., Baskaran, D., \& Li, T. G.~F. 2011, Phys. Rev. D,
  83, 023005

\bibitem[{Zheng {et~al.}(2016)Zheng, Ding, Biesiada, Cao, \& Zhu}]{Zheng2016}
Zheng, X., Ding, X., Biesiada, M., Cao, S., \& Zhu, Z. 2016, Astrophys. J.,
  825, 17

\bibitem[{Zhu {et~al.}(2022)Zhu, Xie, Hu, Liu, Li, Napolitano, Tang, Zhang, \&
  Mei}]{Zhu2022}
Zhu, L.-G., Xie, L.-H., Hu, Y.-M., {et~al.} 2022, Sci. China Phys. Mech.
  Astron., 65, 259811

\end{thebibliography}

\end{document}